\documentclass[10pt,a4paper]{article}

\usepackage[margin=1.8cm]{geometry}		
\usepackage{setspace}

\usepackage{siunitx}

\usepackage{amsmath}
\usepackage{amsfonts}
\usepackage{amssymb}
\usepackage{gensymb}
\usepackage{booktabs}
\usepackage{xspace}
\usepackage{csquotes}
\usepackage{xifthen}
\usepackage{multirow}
\usepackage{color,xcolor}
\usepackage{tikz}
\usepackage[hyphens]{url}
\usepackage[hidelinks, colorlinks = false]{hyperref}

\usepackage[style=authoryear, natbib=true, sorting=ynt]{biblatex}

\renewbibmacro*{begentry}{}%
\renewbibmacro*{finentry}{\usebibmacro{lblek}\finentry}%
\renewbibmacro{lblek}{}%

\DeclareRefcontext{fin}{sorting=nyt}

\renewbibmacro{lblek}{\label{ref:\thefield{entrykey}}}

\DefineBibliographyStrings{english}{%
  backrefpage = {cf.\ page},
  backrefpages = {cf.\ pages},
}

\newenvironment{srvsection}[2]
{
\section{#1}
\label{sec:#2}
}%
{} 

\defbibheading{bibskip}[\bibname]{%
\smallskip  
}
\defbibheading{bibcsec}[\bibname]{%
\subsection{#1}%
}

\newcommand{\rblock}[4]{%
\subsection{#1}
\label{sec:#2}
#4

\printbibliography[#3, heading=bibskip, title={#1}]
}
\newcommand{\oblock}[2]{%
#2%
\printbibliography[#1, heading=bibskip, title={~}]
}
\newcommand{\xcfsec}[1]{(cf.\ Section~\ref{sec:#1})}
\newcommand{\xinsec}[1]{in Section~\ref{sec:#1}}
\newcommand{\questionmrk}[4]{}

\definecolor{darkgreen}{HTML}{117733}
\newcommand{\qcomment}[1]{}
\newcommand{\qdone}[1]{}

\newcommand{\doneedit}[2]{#2}
\newcommand{\droppededit}[2]{#2}

\newcommand{\editrev}[2]{#2}

\colorlet{prevcolor}{black}
\newcommand{\noscite}[1]{}
\newcommand{\seccite}[1]{#1}
\newcommand{\nosciteTL}[1]{\noscite{\colorlet{prevcolor}{gray}#1\colorlet{prevcolor}{black}}}
\newcommand{\secciteTL}[1]{\seccite{\colorlet{prevcolor}{brown}#1\colorlet{prevcolor}{black}}}

\newcommand{\layoutnarw}[1]{}
\newcommand{\layoutwide}[1]{#1}

\newenvironment{pfigure}{\begin{figure}[p]}{\end{figure}}

\newcommand{\citfig}[1]{\noscite{Fig.}\seccite{Figure}~\ref{#1}}

\usepackage{custom_commands}
\usepackage{figdraw_commands}

  \newcommand{\lineSec}{.55}%
  \newcommand{\baseYSecTab}{-2}%
  \newcommand{\baseYSecGrp}{-5.8}%
  \newcommand{\baseYSecTem}{-5.25}%

  \newcommand{\baseXSecTab}{-8.5}%
  \newcommand{\baseXSecGrp}{-8.5}%
  \newcommand{\baseXSecTem}{2.}%

  \newcommand{\baseYMth}{-2.75}%
  \newcommand{\baseXMth}{3.}%

\title{The Minimum Description Length Principle for Pattern Mining\\A Survey\\{\large[v5]}}
\author{Esther Galbrun\\School of Computing, University of Eastern Finland\\\url{esther.galbrun@uef.fi}}
\date{July 2022}

\begin{document}

\maketitle

\begin{center}
\begin{minipage}{0.8\textwidth}
\secciteTL{%
  This is about the Minimum Description Length (MDL) principle applied to pattern mining.
The length of this description is kept to the minimum.

\medskip
Mining patterns is a core task in data analysis and, beyond issues of efficient enumeration, the selection of patterns constitutes a major challenge.
The \noscite{Minimum Description Length (}MDL\noscite{)} principle, a model selection method grounded in information theory, has been applied to pattern mining with the aim to obtain compact high-quality sets of patterns.
After giving an outline of relevant concepts from information theory and coding, as well as of work on the theory behind the MDL and similar principles, we review MDL-based methods for mining various types of data and patterns.
Finally, we open a discussion on some issues regarding these methods, and highlight currently active related data analysis problems.}
\nosciteTL{%
  \input{abstract_dami}}
\end{minipage}
\end{center}


\begin{refsection}[mdl_survey,mdl_survey_v2,mdl_survey_v3,mdl_survey_v4,mdl_survey_v5,infomap]
\nocite{*}
\section{Introduction}
\label{sec:introduction}

Our aim is to review the development of pattern mining methods based on and inspired from the Minimum Description Length (MDL) principle.
Although this is an unrealistic goal, we strive for completeness in our coverage of these methods.

\nosciteTL{%
\label{sec:background-dm}
Mining frequent patterns is a core task in data mining, and itemsets are probably the most elementary and best studied type of pattern. 
Soon after the introduction of the frequent itemset mining task \citep{agrawal_mining_1993}, it became obvious that beyond issues of efficiency \citep{agrawal_fast_1994, mannila_efficient_1994}, the problem of selecting patterns constituted a major challenge to tackle, lest the analysts drown under the deluge of extracted patterns.

\editrev{REPLACED pattern BY itemset IN THE FOLLOWING PARAGRAPH}{}
Various properties and measures have been introduced to select itemsets \citep{webb_efficient_2013,geng_interestingness_2006}. They include identifying representative itemsets \citep{bastide_mining_2000}, using user-defined constraints to filter itemsets \citep{de_raedt_constraint-based_2007,soulet_mining_2011,guns_k-pattern_2013}, considering dependencies between itemsets \citep{jaroszewicz_interestingness_2004,yan_summarizing_2005,tatti_decomposable_2008,mampaey_mining_2010} and trying to evaluate the statistical significance of itemsets \citep{webb_discovering_2007,gionis_assessing_2007,tatti_probably_2010,hamalainen_tutorial_2018}, also looking into alternative approaches to explore the search space \citep{boley_direct_2011,guns_itemset_2011}.  Initially focused on individual itemsets, approaches were later introduced to evaluate itemsets collectively, trying for instance to identify and remove redundancy \citep{gallo_mini_2007}, also in an iterative manner \citep{hanhijarvi_tell_2009,boley_one_2013}. The goal hence moved from mining collections of good patterns to mining good collections of patterns, which is also the main objective when applying the MDL principle.
%


\medskip}

\label{sec:pattern-def}
\editrev{EDITED THE OPENING A BIT}{}
\doneedit{Define what we consider as patterns, divided into substructure and block patterns}{%
Before we go any further, let us explain more precisely what we consider, for the present purpose, to constitute \emph{patterns}.
Patterns are about repetitions. We adopt the point of view that patterns express the repeated presence in the data of particular items, attribute values or other discrete properties. We divide them into two main categories.

Itemsets are strict conjunctive patterns that require the occurrence of all involved items for the pattern to be considered as occurring in a transaction. For a given dataset, it is thus straightforward to determine where an itemset occurs and where it does not, that is, to compute the pattern's support. Vice versa, when it is known that an itemset occurs in a given transaction, no further information is necessary, as it implies that all items must be present. These concepts naturally extend beyond transactional data.
However, such occurrence requirements are fairly strict, and it can be useful to consider more relaxed patterns.
In particular, one might use disjunctions, allowing patterns to express a choice between involved items or attributes. Given a dataset, one can then still straightforwardly determine where a pattern holds. On the other hand, knowing that the pattern occurs no longer unambiguously provides information about which item or attribute is present. More or less additional information is needed, depending for instance on whether it is an inclusive or exclusive disjunction.
We refer to such patterns that express the presence of a specific substructure in the data as \emph{substructure patterns}.

As an alternative way to relax occurrence requirements, patterns might express that selected items or properties are typically present but not all need always occur, for instance by means of density thresholds.
In that case, which data instances belong to the support of the pattern, or vice-versa where each item or property holds, needs to be specified explicitly as it is not directly implied.
Rather than the occurrence of a specific substructure, it is then the homogeneity of repetitions within the area delimited by the selected instances and attributes that is of interest.
Because such patterns can be seen as delineating homogeneous rectangles in the data, we refer to them as \emph{block patterns}.

Furthermore, one is typically looking for a collection of patterns and might impose constraints on the overlap between them, that is, require that the patterns involve disjoint sets of attributes, characterise disjoint sets of instances, or both. In particular, the patterns might be required to form a partition of the data, dividing all instances and all attributes into disjoint subsets.
Such a partitioning requirement is incompatible with a strict occurrence requirement in practice, in the sense that it is not in general possible to identify a collection of substructure patterns that forms a partition of the data. On the other hand, block patterns might be required to form a partition of the data, corresponding roughly to biclustering approaches, or they might be allowed to overlap, as in tiling approaches.

In summary, we adopt a rather broad definition of what constitute patterns, from itemsets to biclusters over discrete data.
However, we stop short of considering clusters more in general as patterns.
Clustering constitutes another important field of data mining beside---and partially overlapping with---pattern mining. There, the goal is to organise data instances into groups such that instances within the same group are similar to each other and dissimilar to instances in other groups. Clustering often handles continuous data, typically relying on a concept of distance. Here, we focus on formalisms and methods that are by nature more discrete.}


\medskip

The reader is expected to be familiar with common pattern mining tasks and techniques, but not necessarily with concepts from information theory and coding, of which we therefore give an overview in Section~\ref{sec:basics}\doneedit{and more}{, before presenting the two main encoding strategies that use respectively substructure and block patterns.} %
\seccite{Background work is covered in Section~\ref{sec:background}. We start with the theory behind the MDL principle and similar principles. Then, we go over a few examples of uses of the principle in the adjacent fields of machine learning and natural language processing. We end with an overview of methods that involve practical compression as a data mining tool or that consider the problem of selecting itemsets.} %
 In Sections~\ref{sec:itemsets}--\ref{sec:sequences}, we turn to the review of MDL-based methods for pattern mining proper. The methods are grouped first by the type of data, then by the type of patterns considered, as outlined in \citfig{fig:structure}. Our focus is on the various encodings designed for these different types of data and patterns, rather than on algorithmic issues related to searching the patterns. In Section~\ref{sec:itemsets}, we start with one of the thickest branches, stemming from the \algKrimp{} algorithm for mining itemsets from transactional data. We continue with itemsets and with other types of patterns for tabular data in Section~\ref{sec:tabular}, followed by graphs and temporal data in Sections~\ref{sec:graphs} and~\ref{sec:sequences}, respectively.
Finally, we consider some discussion points, in Section~\ref{sec:discussion}\seccite{, 
before highlighting related problems that have recently attracted research interest, in Section~\ref{sec:beyond}}. 
\noscite{%
 For further context about the development of the principle and discussion of related problems, please refer to the extended manuscript.\!\footnote{See \urlArXiv{}.}
}
\seccite{%

Sections contain lists of references ordered by publication year, to provide an better overview of the development in the corresponding sub-field.
To keep things simple, even though sometimes related to different sub-fields, each publication is assigned to a single sub-field, that to which it is considered most relevant. For ease of reference, a complete bibliography ordered alphabetically is included at the end. A version of this survey has been published in a peer-reviewed journal.\!\footnote{See \doiDAMI{}.}} %
In addition, the main characteristics and bibliographic details of publications from Sections~\ref{sec:itemsets}--\ref{sec:sequences} have been collected into a searchable table.\!\footnote{See \urlTable{}.}

\begin{figure}[tb]
\centering
\tikzset{
  typ block/.style={rounded corners=2pt, opacity=0.33},
  mth block/.style={rounded corners=10pt, very thick, dtls-gray},
  mth block B/.style={mth block, densely dashed},
  mth block S/.style={mth block, densely dotted},
  mth block l/.style={dtls-gray, inner sep=1pt, anchor=west, align=right, font=\it},
  sec l/.style={inner sep=1pt, anchor=west, align=right},
  sec node/.style={circle, text width=.5cm, fill=black, text=white, inner sep=1pt, align=center, font=\bf},
  label node/.style={rectangle, inner sep=2pt},
  label node T/.style={label node, font=\sc},
  label node L/.style={label node, align=left, anchor=west, xshift=.9em},
  label node R/.style={label node, align=right, anchor=east, xshift=-.9em},
  label node B/.style={label node, align=center, anchor=north, yshift=-1.em},
}

\begin{tikzpicture}[scale=0.8]
  \fill[typ block, blTab!75] (-7.75,0.5) rectangle (.05, 3.55);
  \fill[typ block, blTab!85] (-7.65,2.55) rectangle (-4.3, 3.45);
  \fill[typ block, blTab!85] (-7.65,1.55) rectangle (-4.3, 2.45);
  \fill[typ block, blTab!85] (-4.2,.6) rectangle (-.05, 3.45);

  \fill[typ block, blGrp!75] (-2.1,-1.1) rectangle (3.4, 2);
  \fill[typ block, blTem!75] (.35,-.1) rectangle (6.6, 5.7);
  
  \fill[typ block, blGrp!85] (.45,-0.0) rectangle (3.3, 1.9);
  \fill[typ block, blTem!85] (.45,3.55) rectangle (6.5, 5.6);
  \fill[typ block, blTem!95] (.55,4.5) rectangle (6.4, 5.5);

  \fill[typ block, blTem!85] (.45,3.45) rectangle (6.5, 2.05);
  \fill[typ block, blTem!95] (3.5,3.35) rectangle (6.4, 2.15);

  \draw[mth block B] (1.05,5.45) rectangle (1.95, 2.55);
  \draw[mth block S] (2.05,5.45) rectangle (2.95, 2.55);

  \draw[mth block B] (-.55,1.) rectangle (1.35, 1.9);
  \draw[mth block S] (-.55,0) rectangle (1.35, .9);

  \draw[mth block B] (-2.45,1.55) rectangle (-1.55, 2.45);
  \draw[mth block S] (-3.95,2.55) rectangle (-2.05, 3.45);

\node[label node R] (Ab) at (-4.85,3) {categorical};
\node[sec node] (S55) at (-4.75,3) {\ref{sec:tabular-categorical}};
\node[label node R] (Ac) at (-4.85,2) {numerical};
\node[sec node] (S56) at (-4.75,2) {\ref{sec:tabular-numerical}};

\node[label node T, align=left, anchor=west, text=blTab] (A) at (-7.6,1) {Tabular data};
\node[sec node] (S4) at (-3.5,3) {\ref{sec:itemsets}};
\node[sec node] (S51) at (-2.5,3) {\ref{sec:tabular-binary}};
\node[sec node] (S53) at (-1.5,3) {\ref{sec:tabular-fca}};
\node[label node R] (Aa) at (-2.2,2) {binary};
\node[sec node] (S52) at (-2.,2) {\ref{sec:tabular-tiles}};
\node[sec node] (S54) at (-1.,2) {\ref{sec:tabular-bmf}};

\node[label node T, align=center, text=blGrp] (B) at (1.65,-.55) {Graphs};
\node[sec node] (S61) at (-0.1,1.45) {\ref{sec:graphs-blocks}};
\node[sec node] (S65) at (-0.1,.45) {\ref{sec:graphs-substructures}};
\node[sec node] (S63) at (-1.1,.45) {\ref{sec:graphs-hyperbolic}};
\node[sec node] (S64) at (-1.1,-.55) {\ref{sec:graphs-mapequation}};
\node[sec node] (S67) at (-0.1,-.55) {\ref{sec:graphs-pathways}};

\node[label node L] (Ba) at (1.,.95) {dynamic};
\node[sec node] (S62) at (.9,1.45) {\ref{sec:graphs-blocks-dyn}};
\node[sec node] (S65) at (.9,.45) {\ref{sec:graphs-substructures-dyn}};

\node[label node T, text width=1.8cm, align=center, text=blTem] (C) at (5,1) {Temporal data};
\node[label node L] (Caa) at (3.6,5) {strings};
\node[sec node] (S71) at (1.5,5) {\ref{sec:sequential-seg-haplotype}};
\node[sec node] (S74) at (2.5,5) {\ref{sec:sequential-substring}};
\node[label node L] (Ca) at (3.6,4) {sequences};
\node[sec node] (S72) at (1.5,4) {\ref{sec:sequential-seg-sequences}};
\node[sec node] (S75) at (2.5,4) {\ref{sec:sequential-subsequence}};
\node[sec node] (S77) at (3.5,4) {\ref{sec:sequential-periodicity}};

\node[sec node] (S73) at (1.5,3) {\ref{sec:sequential-seg-timeseries}};
\node[sec node] (S76) at (2.5,3) {\ref{sec:sequential-motifs}};
\node[label node B] (Cb) at (2,3) {time-series};
\node[sec node] (S78) at (4,2.75) {\ref{sec:sequential-trajectories}};
\node[label node L, text width=1.4cm] (Cba) at (4.,2.75) {spatio-temporal};

\draw[mth block B] ({\baseXMth}, {\baseYMth+0.25}) -- ({\baseXMth+.5}, {\baseYMth+0.25});
\node[mth block l] (x) at ({\baseXMth+.75}, {\baseYMth+0.25}) {block patterns};
\draw[mth block S] ({\baseXMth}, {\baseYMth-0.25}) -- ({\baseXMth+.5}, {\baseYMth-0.25});
\node[mth block l] (y) at ({\baseXMth+.75}, {\baseYMth-0.25}) {substructure patterns};

\node[sec l] at (\baseXSecTab,{\baseYSecTab+.1}) {\ref{sec:itemsets}\phantom{.0}\enskip\nameref{sec:itemsets}};
\foreach \l [count=\i] in {%
sec:tabular-binary, sec:tabular-tiles, sec:tabular-fca, sec:tabular-bmf, %
sec:tabular-categorical, sec:tabular-numerical%
}{
    \node[sec l] at (\baseXSecTab,{\baseYSecTab-\lineSec*\i}) {\ref{\l}\enskip\nameref{\l}};
  }
  \foreach \l [count=\i] in {%
    sec:graphs-blocks, sec:graphs-blocks-dyn, sec:graphs-hyperbolic, sec:graphs-mapequation, %
    sec:graphs-substructures, sec:graphs-substructures-dyn, sec:graphs-pathways%
}{
    \node[sec l] at (\baseXSecGrp,{\baseYSecGrp-\lineSec*\i}) {\ref{\l}\enskip\nameref{\l}};
  }
  \foreach \l [count=\i] in {%
 sec:sequential-seg-haplotype, sec:sequential-seg-sequences, sec:sequential-seg-timeseries, %
 sec:sequential-substring, sec:sequential-subsequence, sec:sequential-motifs, %
 sec:sequential-periodicity, sec:sequential-trajectories%
}{
    \node[sec l] at (\baseXSecTem,{\baseYSecTem-\lineSec*\i}) {\ref{\l}\enskip\nameref{\l}};
  }
\end{tikzpicture}
\caption{Organisation of Sections~\ref{sec:itemsets}--\ref{sec:sequences}. MDL-based methods for pattern mining are grouped first by the type of data (tabular, graph or temporal), then by the type of patterns and strategies considered (blocks or substructures). Numbers refer to sections. The three main data types and their subtypes are represented by coloured shapes. Simple unlabelled graphs can be represented as binary matrices. Thus, some methods designed for binary data can be applied to them, and vice versa, some graph-mining methods in effect process binary data. The corresponding sections are therefore represented as lying at the intersection between binary and graph data.
Dashed and dotted lines are used to group methods associated to the two main strategies \xcfsec{basics-strategies}. \emph{Block-based} strategies are used to mine block patterns (also tiles and segments) that group together elements that are similarly distributed. On the other hand, \emph{dictionary-based} strategies are used to mine substructure patterns (also motifs and episodes) that capture specific arrangements and co-occurrences between elements. \qdone{split section 5.1, added Equation Map section}}
\label{fig:structure}
\end{figure}



\section{The basics in a nutshell}
\label{sec:basics}

The Minimum Description Length (MDL) principle is a model selection method grounded in information theory.
\noscite{%
  
\doneedit{moved a few references from the background section}{%
The \citeyear{shannon_mathematical_1948} article by \citeauthor{shannon_mathematical_1948} is widely seen as the birth certificate of information theory, whereas the introduction of the Minimum Description Length principle can be dated back to the seminal paper by \citeauthor{rissanen_modeling_1978} in \citeyear{rissanen_modeling_1978}.
The textbooks by \citet{stone_information_2013} and by \citet{cover_elements_2012} provide respectively an accessible introduction and a more detailed account of information theory and related concepts, while the textbook by \citet{grunwald_minimum_2007} is often regarded as the major reference about the MDL principle.}

The MDL principle is maybe the most popular among several similar principles.}
It can be seen as a practical variant of the Kolmogorov complexity, according to which the complexity of an object, for instance a dataset, is the length of the shortest computer program that outputs it.
The idea is that regularities in the object can be exploited to make the program more concise.
For example, the string that consists of a thousand repetitions of the word \texttt{baobab} can be produced with the following short program: \texttt{for i in range(1000): print(`baobab')}. To output instead a string of the same length where the three letters appear in a completely random manner, we have no choice but to embed the string in the program, resulting in a much longer program. The second string is thus considered to be more complex than the first.

However, the Kolmogorov complexity cannot be computed in general. The MDL principle makes it practical by considering more restricted description methods rather than a universal programming language.
Specifically, when working with the MDL principle, a class of models of the data is considered and the model that allows to most concisely represent the data is selected.

\subsection{Encoding data}
\label{sec:basics-encoding}
Clearly, this is akin to data encoding, where the aim is to map the input object to a sequence, referred to as its \emph{description}, from which the original object can then be reconstructed.
In practical scenarios of data encoding, the aim might be to efficiently and reliably either store the object on a storage medium or transmit it over a communication channel.
The system of rules that dictates how the object is converted into the description and back is called a \emph{code}. The processes of mapping the object to its description and of reconstructing it are called \emph{encoding} and \emph{decoding}, respectively.
The considered storage or channel is typically binary, meaning that the object is mapped to a binary sequence, i.e.\ over the alphabet $\{0, 1\}$, whose length is hence measured in bits.
There has been a lot of studies about communication through noisy channels, that is, when errors might be introduced into the transmitted sequence, and how to recover from it, but this is not of interest to us.
Instead of noise-resistant codes, we focus purely on data compression, on obtaining compact descriptions.
In general, data compression can be either lossless or lossy, depending whether the source object can be reconstructed exactly or only approximately.

Typically, encoding an object means mapping its elementary parts to binary codewords and concatenating them. 
Care must be taken to ensure that the resulting bit-string is decodable, that is, that it can be broken down back into the original codewords from which the parts can be recovered.
For instance, imagine the data consist of the outcome of five throws of a pair of dice, i.e.\ a list of five integers in the interval $[1, 12]$.
If we simply turn the values into their binary representations and concatenate them, we might not be able to reconstruct the list of integers.
For example, there is no way to tell whether \texttt{1110110110} stands for \texttt{11~10~1~10~110}, i.e.\ $\langle3, 2, 1, 2, 6\rangle$, or for \texttt{1~110~1~101~10}, i.e.\ $\langle1, 6, 1, 5, 2\rangle$. To avoid this kind of confusion, we want our code to be such that there is a single unambiguous way to split an encoded string into codewords. 
One strategy is to use separators. For instance, we might represent each integer by as many \texttt{1}s, separated by \texttt{0}s, so that $\langle3, 2, 1, 2, 6\rangle$ becomes \texttt{1110110101101111110}.

More in general, this is where the \emph{prefix-free property} becomes very useful. A \emph{prefix-free code} (also confusingly known as prefix code or instantaneous code) is such that no extension of a codeword can itself be a codeword.

\medskip
\textbf{\emph{Fixed-length codes}} (a.k.a.\ \emph{uniform codes}), that assign distinct codewords of the same length to every symbol in the input alphabet, clearly satisfy the \emph{prefix-free property}. For an input alphabet consisting of $n$ distinct symbols, the codewords must be of length $\lceil \log_2(n) \rceil$. Such a code minimises the worst-case codeword length, that is, the longest codeword is as short as possible. With such a code, every symbol is worth the same. Therefore it is a good option for pointing out an element among canonically ordered options under a uniform distribution, without a priori bias.
For example, the sequence \texttt{baeaecdaeeccbc} over the five-letter alphabet $\langle\texttt{a}, \texttt{b}, \texttt{c}, \texttt{d}, \texttt{e}\rangle$ might be encoded in $42$ bits, as \[\texttt{001~000~100~000~100~010~011~000~100~100~010~010~001~010}\;.\]
\qdone{example}

When symbols are not uniformly distributed, using codewords of \emph{different lengths} can result in codes that are more efficient on average.
There are only so many short codewords available, so one needs to choose wisely what they are used for. 
Intuitively, symbols that are more frequent should be assigned shorter codewords.
The \emph{Kraft--McMillan inequality} (also known simply as Kraft inequality) gives a necessary and sufficient condition for the existence of a prefix-free code. Specifically, it states that, for a finite input alphabet $\mathcal{A}$, the codeword lengths for any prefix-free code $C$ must satisfy \[\sum_{x \in \mathcal{A}} 2^{-L_C(x)} \leq 1\;,\]
where $L_C(x)$ denotes the length of the codeword assigned by $C$ to symbol~$x$.
Vice versa, given codeword lengths satisfying this inequality, there exists a prefix-free code with these codeword lengths.
Furthermore, if $P$ is a probability distribution over a discrete input alphabet $\mathcal{A}$, there exists a prefix-free code $C$ such that for all $x \in \mathcal{A}$, $L_C(x) = \lceil -\log_2\big(P(x)\big) \rceil$.
A pair of related techniques to construct such a varying-length prefix-free code is commonly referred to as \textbf{\emph{Shannon--Fano code}}.
Moreover, given an input alphabet $\mathcal{A}$ where each symbol $x_i$ has an associated weight $w_i$ that might represent, in particular, its frequency of occurrence, a code $C$ is optimal if for any other code $C'$ we have
\[\sum_{x_i \in \mathcal{A}} w_i L_C(x_i) \leq \sum_{x_i \in \mathcal{A}} w_i L_{C'}(x_i)\;.\]
\emph{Huffman's algorithm} is an ingenious simple algorithm allowing to construct an optimal prefix-free code.
Considering the example sequence \texttt{baeaecdaeeccbc} again, Huffman's algorithm would assign shorter codewords to the more frequent letters \texttt{a}, \texttt{c} and \texttt{e}, while ensuring that the prefix-free property is satisfied, allowing for instance to encode the sequence in just $31$ bits, as \[\texttt{011~00~11~00~11~10~010~00~11~11~10~10~011~10}\;.\]
\qdone{example}

\qdone{Prequential coding, two-part vs.\ one-part codes}
Note that to use such a code, the alphabet and the associated probability distribution must be shared by the sender and the receiver.
\editrev{ADDED}{Indeed, it is not enough that the receiver is able to recover the transmitted codewords, the receiver must also know which symbol is represented by each codeword.}
In order to transmit a sequence of symbols, a simple way to proceed is therefore to first transmit the information about the distribution, then to transmit the actual sequence using Shannon--Fano coding, resulting in a \textbf{\emph{two-part code}}. For example, we would first need to transmit the occurrence counts of the five letters, or equivalently their assigned codewords, according to some agreed protocol, before sending the actual encoded $31$-bit sequence.

\medskip
Relying on a sequential prediction, or \textbf{\emph{prequential}}, strategy provides an alternative for coding that does not require the probability distribution to be known a priori. Simply put, the idea is to start with some initial probability distribution over the alphabet, e.g.\ uniform, and at each step encode the next element using a prefix-free code based on the current distribution, then update the probability distribution to account for this occurrence.
These predictive plug-in codes have useful properties. In particular, the total code length does not depend on the order in which the elements are encoded, and is within a constant factor of the optimal.
\qdone{example}

Let us consider the sequence \texttt{baeaecdaeeccbc} as an example once more. We might start with occurrence counts initialised to one for all five letters, and run Huffman's algorithm to assign them codewords accordingly. \editrev{ADDED}{Having agreed on this protocol, the sender and the receiver obtain the same initial codewords, without the sender first having to explicitly share the information about the distribution.}
\editrev{SEPARATE sender AND receiver INSTEAD OF we}{}
The sender first transmits the codeword for letter \texttt{b}, which the receiver can correctly decode using the initial code. Both sides increment their occurrence count of \texttt{b} by one, and update their codewords by running Huffman's algorithm again. Note that at this point, \texttt{b} has the highest occurrence count and will thus be assigned a short codeword ($2$ bits). Next, the sender transmits the new codeword for \texttt{a}, which the receiver is able to correctly decode. The occurrence counts are incremented and the codewords updated, on both sides in parallel. And so on, until the entire sequence has been encoded. \editrev{ADDED}{Crucially, as they apply the same updating protocol in parallel, relying only on information shared so far, sender and receiver maintain identical codes at every step.} Later in the process, the occurrence counts of the letters approach their overall frequencies and the lengths of the assigned codewords converge towards those obtained using Shannon--Fano coding with prior knowledge of the distribution.

\qdone{one-part codes}
More recent advances in MDL and model selection theory introduced Bayesian and normalised maximum likelihood (NML) codes. Like prequential codes, both are \textbf{\emph{one-part codes}}. In contrast to \emph{crude codes}, such \emph{refined codes} remove the need to explicitly encode the parameters of the distribution, thereby avoiding the associated bias, and have useful properties, including optimality guarantees.
In particular, the term \emph{universal} is commonly used to refer to a code that, to put it simply, performs essentially as well as the best-fitting code for the input, for any possible input. 
This use of \emph{universal} should not be confused with another common use, referring to codes for representing integers, which we present next.
On the downside, refined one-part codes are not easily explained in terms of practical encoding and are not always computationally feasible.

\medskip
Finally, a \textbf{\emph{universal code for integers}} is a prefix-free code that maps non-negative integers to binary codewords. Such a code can be used to encode positive integers when the upper-bound cannot be determined a priori. \emph{Elias codes}, which come in three variants---namely Elias gamma code, Elias delta code, and Elias omega code---are universal codes commonly used for this purpose.
For instance, an integer $x \geq 1$ is encoded using the Elias gamma code as $\lfloor \log_2(x) \rfloor$ zeros followed by the binary representation of $x$. These codes penalise large values, since larger integers are assigned longer codewords.

\subsection{Applying the MDL principle in pattern mining}
\label{sec:basics-pm}
It is important to note that when applying the MDL principle, \textbf{compression is used as a tool to compare models}, rather than as an end in itself. 
In other words, we do not care about actual descriptions, only about their lengths. Furthermore, we do not care about the absolute magnitude of the description achieved with a particular model as much as compared to those achieved with other candidate models. This has a few important consequences.

First, as the code does not need to be usable in practice, the requirement of integer code lengths can be lifted, allowing for finer comparisons.
In particular, this means that for a discrete input alphabet $\mathcal{A}$ with probability distribution $P$, the most reasonable choice for assigning codewords is a prefix-free code $C$ such that for all $x \in \mathcal{A}$, $L_C(x) = -\log_2\big(P(x)\big)$. This is often referred to as \emph{Shannon--Fano coding}, although no underlying practical code is actually required, or even considered.
Note that the entropy of $P$ corresponds to the expected number of bits needed to encode in this way an outcome generated by $P$.

Second, elements that are necessary in order to make the encoding system work but are common to all candidate models might be left out, since they play no role in the comparison.
Third, to ensure a fair comparison, only lossless codes are considered. Indeed, comparing models on the basis of their associated description lengths would be meaningless if we are unable to disentangle the savings resulting from a better ability to exploit the structure of the data, versus from increased distortion in the reconstruction. In some cases, this simply means that corrections that must be applied to the decoded data in order to achieve perfect reconstruction are considered as part of the description.

\medskip
\qdone{one-part might be used for components}
Though one-part codes might be used for components, at the highest level most proposed approaches use the \emph{crude \textbf{two-part MDL}}, which requires to first explicitly describe the model $M$, then describe the data using that model, rather than \emph{refined one-part MDL}, which corresponds to using the entire model class $\mathcal{M}$ to describe the data. That is because the aim is not just to know how much the data can be compressed, but how that compression is achieved, by identifying the associated model. 
The overall description length is the sum of the lengths of the two parts, so the best model $M \in \mathcal{M}$ to explain the data $D$ is the one that minimises $L(M,D) =  L(M) + L(D\mid{}M)$, where $L$ denotes description lengths in bits, as above.
The two parts can be understood as capturing respectively the \textbf{complexity of the model} and the \textbf{fit of the model to the data}, and the MDL principle as a way to strike a balance between the two.

\qdone{Swapped next two paragraphs}
One can also view this from \textbf{the perspective of probabilistic modeling}. Consider a family of probability distributions parameterised by some set $\Theta$, that is, $\mathcal{M} = \{ p_{\theta}, \theta \in \Theta\}$, where each $p_\theta$ assigns probabilities to datasets. In addition, consider a distribution $\pi$ over the elements of $\mathcal{M}$.
In this context, in accordance with the Bayesian framework, the best explanation for a dataset $D$ is the element of $\mathcal{M}$ minimising \[ -\log \pi(\theta) - \log p_{\theta}(D)\,,\] where the two terms are the prior and the negative log likelihood of the data, respectively. When using a uniform prior, this means selecting the maximum likelihood estimator.
Since codes can be associated to probability distributions, we can see the connection to the two-part MDL formulation above, where the term representing the description length of the model, $L(M)$, corresponds to the prior, and the term representing the description length of the data given the model, $L(D\mid{}M)$, corresponds to the likelihood.

\doneedit{BIC}{%
The \emph{Bayesian Information Criterion (BIC)} is a closely related model selection method that aims to find a balance between the fit of the model, measured in terms of the likelihood function, and the complexity of the model, measured in terms of the number of parameters $k$. Denoting as $n$ the number of data points in $D$, it can be written as $k \log n - 2 \log p_{\theta}(D)$.}

\medskip
When applying the MDL principle to a pattern mining task, the models considered consist of patterns, capturing structure and regularities in the data.
Depending on the type of data and the application at hand, one must decide what kind of patterns are relevant, i.e.\ \textbf{\textit{(i)} a pattern language must be chosen}.
The model class $\mathcal{M}$ then consists of all possible subsets of patterns from the chosen pattern language.
Note that we generally use the term \emph{model} to refer to a specific collection of patterns and the single corresponding probability distribution over the data, whereas from the statistical modeling perspective, \emph{model} typically refers to a family of probability distributions.
Next, \textbf{\textit{(ii)} an encoding scheme must be designed}, i.e.\ a mechanism must be engineered to encode patterns of the chosen type and to encode the data by means of such patterns. Finally, \textbf{\textit{(iii)} a search algorithm must be devised} to explore the space of patterns and identify the best set of patterns with respect to the encoding, i.e.\ the set of patterns that results in the shortest description length. 

\subsection{Dictionary-based vs.\ block-based strategies}
\label{sec:basics-strategies}

The crude two-part MDL requires to explicitly specify the model (the $L(M)$ part). This means designing \droppededit{encoding protocol, it's about deciding what needs to be transmitted? in what order? using what code? not just the latter}{an ad-hoc encoding scheme} to describe the patterns. This is both a blessing and a curse, because it gives leeway to introduce some bias, i.e.\ penalise properties of patterns that are deemed less interesting or useful than others by making them more costly in terms of code length. But it can involve some difficult design choices and lead to accusations of ``putting your fingers on the scale''. 

\doneedit{for some pattern languages, of the strict occurrence type}{When considering substructure patterns, }encoding the data using the model (the $L(D\mid{}M)$ part) can be fairly straightforward, simply replacing occurrences of the patterns in the data by their associated codewords.
Knowing how many times each pattern $X$ is used in the description of the data $D$ with model $M$, denoted as $\usage_{D, M}(X)$, $X$ can be assigned a codeword of length
\[L(X) = -\log_2 \big(\frac{\usage_{D, M}(X)}{\sum_{Y \in M} \usage_{D, M}(Y)} \big)\] using Shannon--Fano coding.
The design choice of how to cover the data with a given set of patterns, dealing with possible overlaps between patterns in $M$, determines where each pattern is used, defining $\usage_{D, M}(X)$.
The requirement that the encoding be lossless means that elementary patterns, e.g.\ singleton itemsets, must be included in the model, to ensure complete coverage. In this case, encoding the model (the $L(M)$ part) consists in providing the mapping between patterns and codewords, typically referred to as the code table. That is, for each pattern in turn, \droppededit{explain this more?}{describing it} and indicating its associated codeword. Such a \emph{code table} or \textbf{\emph{dictionary-based} strategy}, which corresponds to frequent pattern mining approaches, is a common way to proceed.

An alternative strategy is to divide the data into blocks, \doneedit{other types of patterns which divide the data into blocks/ homogeneous possibly overlapping areas}{that might or might not be allowed to overlap}, each of which is associated with a specific probability distribution over the possible values and should be as homogeneous as possible. The values within each block are then encoded using a code optimised for the corresponding distribution. In that case, encoding the model consists in indicating the \droppededit{how this is done depends on whether overlap is permitted}{limits of each block} and the associated probability distribution. Such a \textbf{\emph{block-based} strategy} corresponds to \doneedit{put this in a different way, esp.\ as it included approaches that allow overlap, which are clearly distinct from clustering}{segmentation, biclustering and tiling approaches}.

\begin{figure}[tbp]
  \OrgData{}
\caption{A toy binary dataset, with six columns (i.e.\ items) denoted $A$--$F$,  nine rows (i.e.\ transactions) denoted $(1)$--$(9)$, containing twenty-four ones (i.e.\ positive entries or item occurrences) represented as black squares.}
\label{fig:data-binary-ex}
\end{figure}


\qdone{Added explanation and illustrative examples}
\droppededit{Explain how the two strategies correspond to groups of pattern types, but cut across data types?}{}
These two main strategies, \emph{dictionary-based} and \emph{block-based}, that use respectively substructure and block patterns, are an important distinguishing factor that we use to categorise approaches, as depicted in \citfig{fig:structure} with dotted and dashed lines, respectively.
We further illustrate and contrast these strategies on a toy binary dataset shown in \citfig{fig:data-binary-ex}. The dataset has six columns (i.e.\ items) denoted $A$--$F$,  nine rows (i.e.\ transactions) denoted $(1)$--$(9)$, and contains twenty-four ones (i.e.\ positive entries or item occurrences) represented as black squares. In particular, the approaches delineated here to illustrate the two strategies are based on the work of \citet{siebes_item_2006} and \citet{smets_slim_2012} \xcfsec{itemsets}, on one hand, and of \droppededit{this case is a partitioning case, why this choice? other end of the spectrum, allows to show how to exploit dependencies between the pieces, e.g.\ a column can belong in only one group, bridges to examples on other data types}{}\citet{chakrabarti_fully_2004} \xcfsec{tabular-tiles}, on the other hand.

\begin{figure}[tbp]
\begin{tabular}{@{}c@{}c@{}c@{}}
\multicolumn{3}{l}{$L(M,D) =$} \\[-1em]
  $L(M)$ & $+$ & $L(D\mid{}M)$ \\
  \hspace{.52\textwidth} & & \hspace{.42\textwidth} \\[-.5em]
  \multicolumn{3}{l}{\textit{i)} The simplest model, with all singleton itemsets, a.k.a.\ \emph{standard code table (ST)}.} \\
  \SlimStandardCT{} & \hspace*{1.6em} & \SlimStandardData{} \\[-1em]
\end{tabular}

\begin{minipage}{.45\textwidth}
\begin{align*}
  \phantom{-2 \cdot \log_2(6/24)}\, -\log_2(6/24)&-\log_2(6/24)\\
  -\log_2(4/24)&-\log_2(4/24)\\
  -\log_2(4/24)&-\log_2(4/24)\\
  -\log_2(4/24)&-\log_2(4/24)\\
  -\log_2(3/24)&-\log_2(3/24)\\
   \underbrace{-\log_2(3/24)}_{\text{itemset}}&\underbrace{-\log_2(3/24)}_{\text{codeword}}\\[.5em]
  =\quad& 31.510
\end{align*}
\end{minipage}\hfill
\begin{minipage}{.34\textwidth}
\begin{align*}
  -6 \cdot \log_2(6/24)&\\
  -4 \cdot \log_2(4/24)&\\
  -4 \cdot \log_2(4/24)&\\
  -1 \cdot \log_2(4/24)&\\
  -3 \cdot \log_2(3/24)&\\
  -3 \cdot \log_2(3/24)&\\[-1.2em]
  &\hspace{-8.5em}\underbrace{\phantom{-3 \cdot \log_2(3/24)}}_{\text{listing itemsets occurrences}}\\[.5em]
  +\quad 61.020 &\quad=\quad 92.530
\end{align*}
\end{minipage}
\vspace{.5em}

\begin{tabular}{@{}c@{}c@{}c@{}}
  \hspace{.52\textwidth} & & \hspace{.42\textwidth} \\[-.5em]
  \multicolumn{3}{l}{\textit{ii)} A non-trivial code table (CT).} \\
  \SlimExCT{} & \hspace*{1.8em} & \SlimExData{}\\[-1em]
\end{tabular}

\begin{minipage}{.45\textwidth}
\begin{align*}
  -\log_2(4/24) -2 \cdot \log_2(3/24)&-\log_2(3/14)\\
  -\log_2(6/24) -\log_2(4/24)&-\log_2(4/14)\\
  -\log_2(6/24) &-\log_2(2/14)\\
  -\log_2(4/24) &-\log_2(1/14)\\
  -\log_2(4/24) &-\log_2(4/14)\\[-1.2em]
  \underbrace{\phantom{-\log_2(4/24) -2 \cdot \log_2(3/24)}}_{\text{itemset}}&\underbrace{\phantom{-\log_2(4/14)}}_{\text{codeword}}\\[.5em]
  =\quad& 32.790
\end{align*}
\end{minipage}\hfill
\begin{minipage}{.34\textwidth}
\begin{align*}
  -3 \cdot \log_2(3/14)&\\
  -4 \cdot \log_2(4/14)&\\
  -2 \cdot \log_2(2/14)&\\
  -1 \cdot \log_2(1/14)&\\
  -4 \cdot \log_2(4/14)&\\[-1.2em]
  &\hspace{-8.5em}\underbrace{\phantom{-4 \cdot \log_2(4/14)}}_{\text{listing itemsets occurrences}}\\[.5em]
  +\quad 30.543 &\quad=\quad 63.333
\end{align*}
\end{minipage}
\caption{Dictionary-based strategy, examples on the toy binary dataset of \citfig{fig:data-binary-ex}. The model $M$ consists of a code table associating patterns, here itemsets, to codewords (left). The data $D$ is encoded using the model by replacing occurrences of the itemsets by the associated codewords (right). Coloured blocks represent prefix-free codewords assigned to items (green) and itemsets (blue), their width is proportional to the code length.}
\label{fig:dictionary-binary}
\end{figure}


\medskip
Examples of \textbf{modeling the toy binary dataset using a dictionary-based approach} are provided in \citfig{fig:dictionary-binary}. The simplest model is considered first \textit{(i)}, which contains as its patterns all singleton itemsets from the dataset and is often referred to as the \emph{standard code table (ST)}.
A non-trivial code table is considered next \textit{(ii)}.
In both cases, the model \doneedit{encoded as/represented by?}{is represented as a code table }associating patterns, here itemsets, to codewords. Encoding the model means encoding each pair. On one hand, \droppededit{an example of encoding protocol}{itemsets are specified} by listing the items they contain. This uses Shannon--Fano coding over the alphabet of items associated to their frequency of occurrence in the data, which is assumed to be shared knowledge. On the other hand, codewords are assigned using Shannon--Fano coding over the alphabet of itemsets included in the code table, associated to their usage.

In \citfig{fig:dictionary-binary}, the prefix-free codewords assigned to items and to itemsets are represented by coloured blocks in shades of green and blue, respectively. The width of a block is proportional to the length of the represented codeword.
For instance, the first row of the non-trivial code table (bottom-left quadrant of \citfig{fig:dictionary-binary}) specifies the first itemset in the code table, in this case $ADE$, listing the three items that constitute it, $A$, $D$ and $E$, using the items' codewords (green blocks), then provides the codeword assigned to $ADE$ by the code based on the usage of itemsets selected in this model (blue block).

Note that for the standard code table, these two prefix-free codes are the same because the standard code table consists of all singleton items and their usage is precisely their frequency of occurrence in the data. Therefore, all codewords in the top half of \citfig{fig:dictionary-binary} are represented as green blocks. For the same reason, the length of the codeword associated to item $x$ by the first code, which does not depend on the code table, is denoted as $L_{ST}(x)$, whereas the length of the codeword associated to itemset $X$ by the second code, which is different for different code tables, is denoted as $L_{CT}(X)$.

Then, encoding the data using the model simply means replacing itemset occurrences by the corresponding codewords.

In summary, the overall description length can be computed as
\begin{equation}\begin{array}{rc@{\;}c@{\;}c}
L(M,D) =&  L(M) & + & L(D\mid{}M) \\[.5em]
 =& \mathlarger{\sum}_{X \in M} \underbrace{\sum_{x \in X} L_{ST}(x)}_{\text{itemset}} + \underbrace{\phantom{\sum_{x \in X}}\hspace{-1.5em}L_{CT}(X)}_{\text{codeword}} & + &  \underbrace{\sum_{X \in M} \usage(X) \cdot L_{CT}(X)}_{\text{listing itemsets occurrences}}\;.
\end{array}\end{equation}

In this example, the standard code table and the non-trivial code table yield an overall description length of $92.530$ bits and $63.333$ bits, respectively. By identifying items that occur frequently together, the latter results in a shorter description length, and one can therefore conclude that it constitutes a better model for the dataset, according to the MDL criterion.

\qdone{discuss about the use of prequential coding}
A similar approach can be built by replacing static Shannon--Fano coding with dynamic prequential coding. In that case, encoding the model only requires to list the itemsets, i.e.\ only the first column of the code table is needed. Then, the data is encoded by enumerating the itemset occurrences using prequential coding \xcfsec{basics-encoding}. In practise, the process of updating the codewords does not actually need to be carried through. Luckily, the description length of the data can be calculated with a (not so simple) formula that does not depend on the order in which the itemsets are transmitted \citep[see][]{budhathoki_difference_2015}.

\begin{figure}[tbp]
\begin{tabular}{@{\hspace{0.15\textwidth}}c@{\hspace{0.05\textwidth}}c@{\hspace{0.05\textwidth}}c@{\hspace{0.15\textwidth}}}
\multicolumn{3}{l}{$L(M,D) =$} \\[-1em]
  $L(M)$ & $+$ & $L(D\mid{}M)$ \\
 \hspace{.35\textwidth} & & \hspace{.35\textwidth} \\[-.5em]
\multicolumn{3}{l}{\textit{i)} The simplest model, with a single block.} \\
  \BlocksSingleModel{} & \phantom{$+$} & \BlocksSingleData{} \\[-1.5em]
\end{tabular}

\begin{minipage}[b]{.6\textwidth}
\begin{align*}
  \underbrace{\CLuniv(9)}_{m=9 \text{ rows}} + \underbrace{9 \cdot \log_2(9)}_{<1,2,3,4,5,6,7,8,9>} &+ 
  \underbrace{\CLuniv(6)}_{n=6 \text{ cols}} + \underbrace{6 \cdot \log_2(6)}_{<1,2,3,4,5,6>} \\
  + \underbrace{\CLuniv(1)}_{k=1 \text{ row group}} + \underbrace{0}_{(<9>)} &+
  \underbrace{\CLuniv(1)}_{l=1 \text{ col group}} + \underbrace{0}_{(<6>)} \\
  + \underbrace{\log_2(9 \cdot 6 +1)}_{24 \text{ ones in } B_{1,1}} \\[.25em]
=\quad& 64.686
\end{align*}
\end{minipage}
\begin{minipage}[b]{.34\textwidth}
\begin{align*}
  \underbrace{- 30 \cdot \log_2(30/54) - 24 \cdot \log_2(24/54)}_{B_{1,1}: <0,1,0,\dots,0,1,1,0,0,0>}\\[.25em]
  +\quad 53.518 \quad=\quad 118.204
\end{align*}
\end{minipage}

\begin{tabular}{@{\hspace{0.15\textwidth}}c@{\hspace{0.05\textwidth}}c@{\hspace{0.05\textwidth}}c@{\hspace{0.15\textwidth}}}
  \hspace{.35\textwidth} & & \hspace{.35\textwidth} \\[-.5em]
\multicolumn{3}{l}{\textit{ii)} A non-trivial model partitioning the dataset into six blocks.} \\
  \BlocksExModel{} & \phantom{$+$} &  \BlocksExData{} \\[-4em]
\end{tabular}
\begin{minipage}[b]{.6\textwidth}
\begin{align*}
  \underbrace{\CLuniv(9)}_{m=9 \text{ rows}}\hspace{-1mm} + \underbrace{9 \cdot \log_2(9)}_{<7,3,5,9,6,2,4,8,1>} &+ 
  \underbrace{\CLuniv(6)}_{n=6 \text{ cols}}\hspace{-1mm} + \underbrace{6 \cdot \log_2(6)}_{<1,4,5,3,2,6>} \\
  + \hspace{-2mm}\underbrace{\CLuniv(3)}_{k=3 \text{ row groups}}\hspace{-2mm} + \underbrace{\log_2(7) + \log_2(4)}_{<4,3(,2)>} &+
  \hspace{-2mm}\underbrace{\CLuniv(2)}_{l=2 \text{ col groups}}\hspace{-1mm} + \underbrace{\log_2(5)}_{<3(,3)>} \\
  + \underbrace{\log_2(4 \cdot 3 +1)}_{1 \text{ one in } B_{1,1}} + 
    \underbrace{\log_2(4 \cdot 3 +1)}_{10 \text{ ones in } B_{1,2}} &+ 
    \underbrace{\log_2(3 \cdot 3 +1)}_{9 \text{ ones in } B_{2,1}} \\
  + \underbrace{\log_2(3 \cdot 3 +1)}_{2 \text{ ones in } B_{2,2}} + 
    \underbrace{\log_2(2 \cdot 3 +1)}_{0 \text{ one in } B_{3,1}} &+ 
    \underbrace{\log_2(2 \cdot 3 +1)}_{2 \text{ ones in } B_{3,2}} \\[.25em]
=\quad& 88.279
\end{align*}
\end{minipage}
\begin{minipage}[b]{.34\textwidth}
\begin{align*}
  \underbrace{- 11 \cdot \log_2(11/12) - 1 \cdot \log_2(1/12)}_{B_{1,1}: <0,0,0,0\dots0,0,0,0>} \\
  \underbrace{- 2 \cdot \log_2(2/12) - 10 \cdot \log_2(10/12)}_{B_{1,2}: <1,1,1,0\dots1,1,1,0>} \\
  \underbrace{- 0 \cdot \log_2(0/9) - 9 \cdot \log_2(9/9)}_{B_{2,1}: <1,1,1,1,1,1,1,1,1>} \\
  \underbrace{- 7 \cdot \log_2(7/9) - 2 \cdot \log_2(2/9)}_{B_{2,2}: <0,1,1,0,0,0,0,0,0>}\\
  \underbrace{- 6 \cdot \log_2(6/6) - 0 \cdot \log_2(0/6)}_{B_{3,1}: <0,0,0,0,0,0>}\\
  \underbrace{- 4 \cdot \log_2(4/6) - 2 \cdot \log_2(2/6)}_{B_{3,2}: <1,0,0,0,1,0>} \\[.25em]
  +\quad 25.154 \quad=\quad 113.433
\end{align*}
\end{minipage}
\caption{Block-based strategy, examples on the toy binary dataset of \citfig{fig:data-binary-ex}. The model $M$ partitions the dataset $D$ into blocks, each associated to a specific probability distribution over the values (left), which is used to encode the entries (right). More intense shades of orange represent higher probabilities of ones within the corresponding block.}
\label{fig:blocks-binary}
\end{figure}


\medskip
Examples of \textbf{modeling the toy binary dataset using a block-based approach} are provided in \citfig{fig:blocks-binary}. The simplest model is considered first \textit{(i)}, which consists of a single block. A non-trivial model dividing the dataset into six blocks is considered next \textit{(ii)}.
\doneedit{partitioning case, allows to show how to exploit dependencies between the pieces, e.g.\ a column can belong in only one group}{The approach illustrated here requires the patterns to form a partition of the data. Therefore, any considered model consists of a set of non-overlapping blocks covering the entire dataset and, more specifically, is such that the rows and the columns are divided into disjoint groups. This requirement means that each row and each column belongs to exactly one group, a fact that can be exploited when designing the encoding, as explained below. }
Each block is associated to a specific probability distribution over the values occurring within it. In this case, the values are zero and one, corresponding to a Bernoulli distribution. Encoding the model means specifying the groups of rows and of columns that define the blocks, and the probability associated to each one. In this approach, all blocks are induced by a unique partition of the rows and of the columns, which can be specified by defining an order over the rows (resp.\ columns) and indicating the number of rows (resp.\ columns) in each group. Finally, the number of ones in each block is transmitted, allowing to compute the corresponding probability.
This is achieved through a combination of fixed-length and universal codes.

Next, we delve deeper into the details of the encoding scheme, to illustrate the choices that are involved in its design.
We let
\begin{itemize}
  \item $m$ and $n$ denote respectively the number of rows and columns in the dataset,
  \item $k$ and $l$ denote respectively the number of row and column groups,
  \item $m_i$ and $n_j$ denote respectively the number of rows and columns in block $B_{i,j}$,
  \item $\gamma_v(B_{i,j})$ denote the number of entries in block $B_{i,j}$ equal to $v$, and
  \item $\CLuniv(x)$ denote the MDL optimal universal code length for integer $x$.\!\footnote{$\CLuniv(x)$ is defined for $x > 1$ as $\CLuniv(x) = \log_2(x) + \log_2 \log_2(x) + \dots + \log_2(c_0)$, with the value of $c_0$ set so as to satisfy the Kraft inequality. Specifically, we let $c_0 = 2.86507$.}
  \end{itemize}
\qdone{Added note about actual universal code used in examples $\CLuniv$}
The formula for computing the overall description length is provided in Equation~\ref{eq:blocks}.
The number of rows and the number of columns are transmitted using universal coding, since these values are not bounded a priori, and the order of rows (resp.\ of columns) is then specified by listing row (resp.\ column) identifiers in turn, using a fixed-length code (cf.\ part $(a)$ of Equation~\ref{eq:blocks}).
In fact, this part of the encoding is independent of the model and has a constant length for a given dataset. Therefore, it does not impact the comparison and can be ignored.
The number of row groups $k$ (resp.\ column groups $l$)  could be transmitted using a fixed-length code $\log_2(m)$ (resp.\ $\log_2(n)$), since there cannot be more groups than there are rows (resp.\ columns). However, universal coding is used instead, to favour partitions with fewer groups.
Then, assuming that the numbers of rows and of columns in the groups are sorted and transmitted by decreasing order, upper bounds $m^{*}_i$ and $n^{*}_j$ on $m_i$ and $n_j$, respectively, can be derived given shared knowledge since already transmitted values constrain the remaining ones:
  $m^{*}_i = \big( \sum_{t=i}^{k} m_t \big) - k +i$, for $i = 1,\dots k-1$ and
  $n^{*}_j = \big( \sum_{t=j}^{l} n_t \big) - l +j$, for $j = 1,\dots l-1$.
  
These upper bounds are used to transmit the numbers of rows and of columns in the groups with fixed-length codes (cf.\ part $(b)$ of Equation~\ref{eq:blocks}).
Also the number of ones in each block can be transmitted using a fixed-length code, since it takes value between zero and the number of entries in the block  (cf.\ part $(c)$ of Equation~\ref{eq:blocks}). 

The data is then encoded using the model. This is done by listing the entries in each block with a prefix-free code adjusted to the probability distribution within the block (cf.\ part $(d)$ of Equation~\ref{eq:blocks}).

In summary, the overall description length can be computed as
\begin{equation}\begin{array}{rc@{\;}c@{\;}c@{}c}
L(M,D) =&  L(M) & + & L(D\mid{}M) & \\[.5em]
= & \multicolumn{2}{l}{\underbrace{\CLuniv(m)}_\text{nb.\ rows} + \underbrace{m \cdot \log_2(m)}_\text{ordered rows} +
    \underbrace{\CLuniv(n)}_\text{nb.\ cols} + \underbrace{n \cdot \log_2(n)}_\text{ordered cols}} & & 
\hspace*{-2.5em} \multirow{1}{*}{\tikz{%
\draw [decorate, decoration = {brace, amplitude=5pt, mirror}] (0,0) --  (0,.8) node[pos=0.5,right=5pt,black]{$(a)$};%
}}\\
  & \multicolumn{3}{l}{+ \underbrace{\CLuniv(k)}_\text{nb.\ row groups} + \underbrace{\sum_{i=1}^{k-1} \log_2(m^*_i)}_\text{nb.\ rows in each group}} & %
\hspace*{-2.5em} \multirow{2}{*}{\tikz{%
 \draw [decorate, decoration = {brace, amplitude=5pt, mirror}] (0,0) --  (0,3) node[pos=0.5,right=5pt,black]{$(b)$};%
  }}\\
 & \multicolumn{3}{l}{+ \underbrace{\CLuniv(l)}_\text{nb.\ col groups} + \underbrace{\sum_{j=1}^{l-1} \log_2(n^*_j)}_\text{nb.\ cols in each group}} & \\
 & + \underbrace{\sum_{i=1}^{k} \sum_{j=1}^{l} \log_2(m_i \cdot n_j + 1)}_\text{nb.\ ones in each block} & + & & 
\hspace*{-2.5em} \multirow{1}{*}{\tikz{%
\draw [decorate, decoration = {brace, amplitude=5pt, mirror}] (0,0) --  (0,.8) node[pos=0.5,right=5pt,black]{$(c)$};%
}}\\
\multirow{1}{*}{\tikz{%
\draw [decorate, decoration = {brace, amplitude=5pt}] (0,0) --  (0,.8) node[pos=0.5,left=5pt,black]{$(d)$};%
}} & & \multicolumn{2}{r}{\hspace{-1.8em}\underbrace{\sum_{i=1}^{k} \sum_{j=1}^{l} \sum_{v\in\{0,1\}} -\gamma_v(B_{i,j}) \cdot \log_2\big( \frac{\gamma_v(B_{i,j})}{m_i \cdot n_j} \big)}_\text{listing entries in each block}}\;. 
\end{array}
\label{eq:blocks}
\end{equation}

In \citfig{fig:blocks-binary}, shades of orange are used to represent probability distributions within the blocks, with more intense shades representing higher probabilities of ones. 
In this example, the single-block model and the six-block model yield an overall description length of $118.204$ bits and $113.433$ bits, respectively. By partitioning the dataset into blocks that are particularly dense or particularly sparse, the latter results in a shorter description length, and one can therefore conclude that it constitutes a better model for the dataset, according to the MDL criterion.

\doneedit{Add something about removing the partition constraint and how that affects the encoding?}{%
Removing the partition constraint and allowing overlaps between the block patterns would require to modify the encoding since we could no longer assume that each row and each column belongs to a single group.}

\subsection{Algorithms}
\label{sec:basics-algorithms}

The MDL principle provides a basis for designing a score for patterns, but no way to actually find the best collection of patterns with respect to that score.
The space of candidate patterns, and even more so the space of candidate pattern sets, is typically extremely large, if not infinite, and rarely possesses any useful structure. Hence, exhaustive search is generally infeasible, heuristic search algorithms are employed, and one must be satisfied with finding a good set of patterns rather than an optimal one.
Mainly, algorithms can be divided between
\textit{(i)} approaches that generate a large collection of patterns then (iteratively) select a small set out of it, and
\textit{(ii)} approaches that generate candidates on-the-fly, typically in a levelwise manner, possibly in an anytime fashion.
\qdone{The generate-and-filter search strategy as a simple algo to build a proof-of-concept for a proposed approach, to be replaced by a on-the-fly candidate generation algo}
The former approaches are typically less efficient, since they generate many more candidates than necessary, but constitute a useful basis for building a proof-of-concept.
Because recomputing costs following the addition or removal of a candidate pattern is often prohibitively expensive, an important component of the heuristic search algorithms consists in efficiently and accurately bounding these costs.

\qdone{Explained link between algorithms bottom-up vs. top-down and dictionary vs. block-based strategies}
With a dictionary-based strategy, the simplest model consists of all separate basic elements. The search for candidates can thus start from this model and progressively combine elements.
Vice versa, with a block-based strategy, the simplest model consists of a single block covering the entire dataset. The search for candidates can thus start from this model and progressively split elements.
Intuitively, the two main strategies lend themselves respectively to bottom-up and top-down iterative exploration algorithms. 

\bigskip
In summary, to apply the MDL principle to a pattern mining task, one might proceed in three main steps. First, define the pattern language, deciding what constitutes a structure of interest given the data and application considered. Second, define a suitable encoding scheme, designing a system to encode the patterns and to encode the data using such patterns. Third, design a search algorithm, allowing to identify in the data a collection of patterns that yields a good compression under the chosen encoding scheme.

Our main focus here is on the second step, the design of an encoding scheme for the different types of patterns, which is the core of the MDL methodology and its distinctive ingredient. For the most part, we do not discuss search algorithms and are not concerned by issues of performance, which are more generic aspects of pattern mining methodologies.



\secciteTL{%
\begin{srvsection}{Theoretical and conceptual background}{background}
\qdone{changed title}
  
\oblock{keyword={background}, keyword={information-theory}}{%
The MDL principle is maybe the most popular among several similar principles rooted in information theory.
The \citeyear{shannon_mathematical_1948} article by \citeauthor{shannon_mathematical_1948} is widely seen as the birth certificate of information theory.
The textbooks by \citet{stone_information_2013} and by \citet{cover_elements_2012} provide respectively an accessible introduction and a more detailed account of information theory and related concepts\noscite{~\citep[see also][]{stone_information_2018}}. The textbook of \citet{li_introduction_2019}, on the other hand, focuses on Kolmogorov complexity.
The tutorial by \citet{csiszar_information_2004} covers applications of information theory in statistics and discusses the MDL principle in its last chapter.
}

\rblock{The Minimum Description Length (MDL) principle}{background-mdl}{keyword={background}, keyword={mdl}}{%
  The introduction of the Minimum Description Length principle can be dated back to the seminal paper by \citeauthor{rissanen_modeling_1978} in \citeyear{rissanen_modeling_1978}.
Works collected in \citep{grunwald_advances_2005} present the theoretical foundations of the principle, as well as later advances and applications.
The textbook by \citet{grunwald_minimum_2007} is often regarded as the major reference about the MDL principle.

More recently, \citet{grunwald_minimum_2019} present the MDL principle from the perspective of probabilistic modeling, without resorting to information theory. They review recent theoretical developments, which allow to see MDL as a generalisation of both penalised likelihood and Bayesian approaches.   
\citet{vitanyi_minimum_2000} \citep[also][]{li_computational_1995,vitanyi_minimum_1999} draw also parallels between the Bayesian framework and the MDL principle.
%

\noscite{%
  Over the four decades since its introduction, multiple accounts of the principle and its use for model selection have been given~\citep{rissanen_universal_1983, rissanen_stochastic_1989, barron_minimum_1998, hansen_model_2001, lee_introduction_2001, grunwald_tutorial_2004, rissanen_introduction_2005, grunwald_advances_2005, rissanen_information_2007, de_rooij_luckiness_2011, roos_minimum_2016, vreeken_modern_2019}.}

}

\rblock{Other information-theoretic principles for model selection}{background-other}{keyword={background}, keyword={other}}{%
Other minimisation principles based on information theory, and often closely related to Kolmogorov complexity or its extensions \citep[see e.g.][]{vereshchagin_kolmogorovs_2004}, were being developed around the same time as the MDL principle.
\citet{chaitin_theory_1975} suggested to refer to this emergent field bringing together Shannon's information theory and Turing's computational complexity as Algorithmic Information Theory (AIT).

In 1964, \citeauthor{solomonoff_formal_1964} proposed the concept of algorithmic probability and a theory of inductive inference\noscite{~\citep{solomonoff_formal_1964,solomonoff_formal_1964-2}}. This theory inspired the development of the MDL principle, as well as, independently, the development of the minimal representation criterion by \citeauthor{segen_minimal_1979} in \citeyear{segen_minimal_1979} \citep[also][]{segen_incremental_1989,segen_graph_1990}.

The MDL principle was also partly inspired from the Minimum Message Length (MML) principle, introduced by \citeauthor{wallace_information_1968} in \citeyear{wallace_information_1968} \citep[also][]{wallace_statistical_2005}. The two principles have some important conceptual differences but often lead to similar results. They are discussed and compared by \citet{lanterman_schwarz_2001}, together with other related approaches.
Several applications and discussions of the principles are presented as contributions to the 1996 conference on Information, Statistics and Induction in Science~\citep{dowe_proceedings_1996}.

More recently, \citet{tishby_information_2000} introduced the Information Bottleneck (IB) method~\citep[see also][]{slonim_information_2002}, arguing that it more appropriately focuses on \emph{relevant} information than previously formulated principles.
}

\rblock{General considerations on simplicity, parsimony, and modeling}{background-parsimony}{keyword={background}, keyword={parsimony}}{%
Several authors have contributed to the discussion on conceptual issues related to complexity in modeling.

Among them, \citet{davis_predictive_1996}, and later \citet{robinet_mdlchunker_2011}, examine issues of model selection and parsimony, in relation to human cognition.
\citet{rathmanner_philosophical_2011} propose a rather broad but not overly technical discussion of induction in general and Solomonoff's induction in particular, as well as several associated topics.
\citet{lattimore_no_2013} discuss algorithmic information theory in the context of the \emph{no free lunch theorem}, which, simply put, posits that an algorithm that performs well on particular problems must pay for it with reduced performance on other problems, and \emph{Occam's razor}, a general rule favouring simplicity.
\citet{domingos_role_1999} exposes misconceptions and misuses of Occam's razor in model selection.

\citet{bloem_two_2015} \citep[also][]{bloem_single_2016} suggest to use \emph{sophistication} as an umbrella term for various concepts related to the amount of information contained in data and discuss different approaches that have been proposed for measuring it.

\citet{furnkranz_cognitive_2020} present an insightful investigation into the perception of rule-based models by analysts, highlighting that shorter is not always better. 
\qdone{Added \citep{furnkranz_cognitive_2020}}
}

\rblock{Compression and Data Mining (DM)}{background-compression}{keyword={background}, keyword={compression}}{%
Various approaches using practical data compression as a tool for data mining have been proposed and discussed.

For instance, \citet{keogh_towards_2004} \citep[also][]{keogh_compression-based_2007} present the Compression-based Dissimilarity Measure (CDM), that evaluates the relative gain when compressing two strings concatenated rather than separately. Inspired from Kolmogorov complexity, the measure is used while mining timeseries, to fight against large numbers of parameters in the algorithms.
\qdone{Added Normalized Compression Distance}
Similarly, \citet{cilibrasi_clustering_2005} define a \emph{Normalized Compression Distance}, which they then use for clustering. 

\citet{simovici_minability_2013} \citep[also][]{simovici_compression_2015} proposes to evaluate the presence of patterns using practical data compression. The aim is to detect whether patterns are present, not to find them.

From a more conjectural perspective, \citet{faloutsos_data_2007} argue that the strong connection between data mining, compression and Kolmogorov complexity means there is little hope for a unifying theory of data mining.

The term \emph{compression} as used by \citet{chandola_summarization_2007} is a metaphor rather than a practical tool. The proposed approach for mining patterns from tables of categorical attributes relies on a pair of ad-hoc scores that intuitively measure how much the considered collection of patterns allows to compact the data, and how much information is lost, respectively.
}

\rblock{Some early uses of MDL in Machine Learning (ML)}{background-ml}{keyword={background}, keyword={ml}}{%
\qdone{Added \citep{muggleton_compression_1992}}
The MDL principle has been applied early on in machine learning. Examples include evaluating hypotheses in inductive logic programming \citep{muggleton_compression_1992}, learning Bayesian networks \citep{suzuki_construction_1993}, decision trees \citep{quinlan_inferring_1989,wallace_coding_1993,mehta_mdl-based_1995,robnik-sikonja_pruning_1998}, rules \citep{pfahringer_new_1995}, or other related models \citep{quinlan_minimum_1994,kilpelainen_mdl_1995}, as well as feature engineering \citep{derthick_minimum_1990,derthick_minimal_1991,pfahringer_compression_1995}, supervised discretisation \citep{fayyad_multi-interval_1993}, signal smoothing \citep{pednault_experiments_1989} and segmentation \citep{merhav_minimum_1993,shamir_asymptotically_1999}.
}

\rblock{Some uses of MDL in Natural Language Processing (NLP)}{background-nlp}{keyword={background}, keyword={nlp}}{%
The MDL principle is also used in Natural Language Processing, in text and speech processing, in particular for clustering tasks \citep{li_clustering_1997,li_word_1998}, for segmentation at the level of phrases \citep{kit_goodness_1998}, words \citep{de_marcken_unsupervised_1995,kit_unsupervised_1999,argamon_efficient_2004} or morphemes \citep{brent_discovering_1995,creutz_unsupervised_2002}, and to extract other types of text patterns \citep{li_generalizing_1998,wu_unsupervised_2010}. Some of the works are closely related to the development of methods for mining strings, and for analysing sequential data more generally \xcfsec{sequences}.
}

\rblock{Mining and selecting itemsets}{background-dm}{keyword={background}, keyword={dm}}{%
Mining frequent patterns is a core task in data mining, and itemsets are probably the most elementary and best studied type of pattern. 
Soon after the introduction of the frequent itemset mining task \citep{agrawal_mining_1993}, it became obvious that beyond issues of efficiency \citep{agrawal_fast_1994, mannila_efficient_1994}, the problem of selecting patterns constituted a major challenge to tackle, lest the analysts drown under the deluge of extracted patterns.

\editrev{REPLACED pattern BY itemset IN THE FOLLOWING PARAGRAPH}{}
Various properties and measures have been introduced to select itemsets \citep{webb_efficient_2013,geng_interestingness_2006}. They include identifying representative itemsets \citep{bastide_mining_2000}, using user-defined constraints to filter itemsets \citep{de_raedt_constraint-based_2007,soulet_mining_2011,guns_k-pattern_2013}, considering dependencies between itemsets \citep{jaroszewicz_interestingness_2004,yan_summarizing_2005,tatti_decomposable_2008,mampaey_mining_2010} and trying to evaluate the statistical significance of itemsets \citep{webb_discovering_2007,gionis_assessing_2007,tatti_probably_2010,hamalainen_tutorial_2018}, also looking into alternative approaches to explore the search space \citep{boley_direct_2011,guns_itemset_2011}.  Initially focused on individual itemsets, approaches were later introduced to evaluate itemsets collectively, trying for instance to identify and remove redundancy \citep{gallo_mini_2007}, also in an iterative manner \citep{hanhijarvi_tell_2009,boley_one_2013}. The goal hence moved from mining collections of good patterns to mining good collections of patterns, which is also the main objective when applying the MDL principle.
%


}
\end{srvsection}

}
\begin{srvsection}{Mining itemsets with \algKrimp{} \& Co.}{itemsets}
  
\oblock{keyword={itemset}, keyword={XXX}}{%
A transactional database consists of a collection of sets, called transactions, over a universe of items. The prototypical example for this type of data comes from market-basket analysis, which is also where some of the terminology is borrowed from. Alternatively, a transactional database can be represented as a binary matrix.
\droppededit{Say something about strict occurrence?}{Frequent itemset mining}, that is, finding items that frequently co-occur in a transactional database, is a central task in data mining \xcfsec{background-dm}.
}

\rblock{\algKrimp{}}{itemset-krimp}{keyword={itemset}, keyword={krimp}, keyword={basic}}{%
  The introduction by \citeauthor{siebes_item_2006} in \citeyear{siebes_item_2006} of a MDL-based algorithm for mining and selecting small but high-quality collections of itemsets sparked a productive line of research, including algorithmic improvements, adaptations to different tasks, and various applications of the original algorithm.
%
%
  The algorithm, soon dubbed \algKrimp{} \citep{van_leeuwen_compression_2006}, Dutch for ``to shrink'', is a prime example of a \emph{dictionary-based} strategy \xcfsec{basics-strategies}, illustrated in \citfig{fig:dictionary-binary}.

  Through an evaluation on a classification task, \citet{van_leeuwen_compression_2006} show that the selected itemsets are representative. Specifically, considering a labelled training dataset, \algKrimp{} is applied separately on the transactions associated to each class to mine a code table. A given test transaction can then be encoded using each of the code tables, and assigned to the class that corresponds to the shortest code length.

\noscite{%
The \algKrimp{} algorithm has subsequently been presented with varying depths of technicality \citep{van_leeuwen_identifying_2009,vreeken_krimp_2011,siebes_mdl_2014,van_leeuwen_mining_2014}. It is a corner stone of a couple of doctoral dissertations \citep{vreeken_making_2009,van_leeuwen_patterns_2010} and also contributed to spark a more conceptual discussion \citep{siebes_queries_2012}.%
}
}

\rblock{Algorithmic improvements}{itemset-improve}{keyword={itemset}, keyword={krimp}, keyword={improve}}{%
Works building on \algKrimp{} include several algorithmic improvements.
In particular, the \algSlim{} algorithm of \citet{smets_slim_2012} modifies \algKrimp{} and greedily generates candidates by merging patterns, instead of evaluating candidates from a pre-mined list. The example presented in \citfig{fig:dictionary-binary}\textit{(ii)} \xcfsec{basics-strategies} was actually obtained by running the \algSlim{} algorithm on the toy dataset of \citfig{fig:data-binary-ex}.
\citet{hess_shrimp_2014} propose a data structure similar to the FP-tree to facilitate the recomputation of usages when the \algKrimp{} code table is updated, making the mining algorithm more efficient.
\citet{sampson_widened_2014} apply widening, i.e.\ diversification of the search, to \algKrimp{}.
}

\rblock{Finding differences and anomalies}{itemset-comp}{keyword={itemset}, keyword={comp}}{%

The analysis of differences between databases and the detection of anomalies are derivative tasks that have attracted particular attention.
\citet{vreeken_characterising_2007} use \algKrimp{} to measure differences between two databases, by comparing the length of the description of a 
database obtained with a code table induced on that database versus one induced on the other database. The coverage of individual transactions by the selected itemsets, and the specific code tables obtained are also compared.
The \algname{DiffNorm} algorithm introduced by \citet{budhathoki_difference_2015} aims to encode multiple databases at once without redundancy, and allows to investigate the differences and similarities between the databases by inspecting the obtained code table.
As a major contribution, \citet{budhathoki_difference_2015} improve the encoding by replacing the Shannon--Fano code used in the original  \algKrimp{} by a prequential plug-in code \xcfsec{basics-encoding}.

\citet{smets_odd_2011} use \algKrimp{} for outlier detection by looking at how many bits are needed to encode a transaction. If this number is much larger than expected, the transaction is declared anomalous. In addition, the encodings of transactions can be scrutinised to obtain further insight into how they depart from the rest.
\citet{akoglu_fast_2012} design an algorithm that detects as anomalies transactions having a high encoding cost. Their proposed algorithm mines multiple code tables, rather than a single one in \algKrimp{}, and handles categorical data.
\citet{bertens_beauty_2015} \citep[also][]{bertens_efficiently_2017} propose a method to detect anomalous co-occurrences based on the \algKrimp{}/\algSlim{} code tables.
}

\rblock{Mining rule sets}{itemset-rules}{keyword={itemset}, keyword={rules}}{%
Going beyond itemsets, a closely related task is to mine rules.
\citet{van_leeuwen_association_2015} propose to mine association rules across a two-view transactional dataset, such that one view can be reconstructed from the other, and vice versa.
\citet{fischer_sets_2019} instead consider a unique set of items and aim to mine associations rules that allow to reconstruct the dataset, enabling corrections in order to increase the robustness of the results. They then apply the approach to learn rules about activation patterns in neural networks \citep{fischer_whats_2021}. 
\qdone{\citet{fischer_whats_2021} use it to learn rules about activation patterns in neural networks}

\citet{aoga_finding_2018} present a method to encode a binary label associated to each transaction, using the original transactions and a list of rules, each associated to a probability that the target variable holds true.
\citet{proenca_interpretable_2020} \citep[also][]{proenca_interpretable_2020-1} consider a similar task, but with multiple classes and targeted towards predictive rather than descriptive rules, 
then looking for rules that capture deviating groups of transactions, i.e.\ dealing with the subgroup discovery task \citep{proenca_discovering_2020, proenca_discovering_2021}.
Beside binary attributes, i.e.\ items, \citet{proenca_discovering_2020} also consider nominal and numerical attributes, and aim to predict a single numerical target, modeled using normal distributions, instead of a binary target. \citet{proenca_robust_2021} further extend the approach to more complex nominal and numerical targets by resorting to normalised maximum likelihood (NML) and Bayesian codes, respectively. 
\qdone{\citet{proenca_robust_2021} and more complex targets: nominal attributes, NML encoding of the targets}

Whereas the former two output a set of rules, these latter methods return a list of rules, such that at most one rule, the first valid rule encountered in the list, applies to any given transaction. In all cases, the dataset, or part of it, is assumed to be given and the aim is to reconstruct a target variable, or the rest of the dataset.
}

\rblock{Other adaptations}{itemset-kmore}{keyword={itemset}, keyword={krimp}, keyword={more}}{%
Further work also includes extending the \algKrimp{} approach to more expressive patterns, such as patterns mined from relational databases \citep{koopman_discovering_2008,koopman_characteristic_2009}, and using \algKrimp{} for derivative tasks.

\citet{van_leeuwen_streamkrimp_2008} use \algKrimp{} to detect changes in the distribution of items in a streaming setting. If a code table induced from an earlier part of the stream no longer provides good compression as compared to a code table induced from a more recent part of the stream, it is a signal that the distribution has changed.
\citet{bonchi_characterizing_2011} extend the approach to the probabilistic setting, where the occurrence of an item in a transaction is associated to a probability, aiming to find a collection of itemsets that compress the data well in expectation. 

Given an original database, \citet{vreeken_preserving_2007} use \algKrimp{} to generate a synthetic database similar to the original one, for use in the context of privacy-preserving data mining. The code table is induced on the original database, itemsets are then sampled from it and combined to generate synthetic transactions.
\citet{vreeken_filling_2008} use \algKrimp{} for data completion. In a way, this turns the MDL principle on its head. Starting from an incomplete database, rather than looking for the patterns that compress the data best, the proposed approach looks for the data that is best compressed by the patterns and considers that to be the best completion.
Instead of a single code table, \citet{siebes_structure_2011} look for a collection of code tables, that capture the structure of the dataset at different levels of granularity. Representing a categorical dataset as transactional data by mapping each value of an attribute to a distinct item implies that each transaction contains exactly one item for each categorical attribute, and hence that all transactions have the same length. \citet{siebes_smoothing_2012} consider the problem of smoothing out the small scale structure from such datasets. That is, they aim to replace entries in the data so that its large scale structure is maintained but it can be compressed better.   
}

\rblock{Applications}{itemset-applications}{keyword={itemset}, keyword={applications}}{%
The \algKrimp{} algorithm has also been employed to tackle problems in different domains, including clustering tagged media \citep{van_leeuwen_compressing_2009}, summarising text \citep{vanetik_query-based_2017,vanetik_drim_2018}, detecting malware \citep{asadi_mdl-based_2019,asadi_towards_2019} and analysing the Semantic Web \citep{bobed_data-driven_2019}.
}
\end{srvsection}

\begin{srvsection}{Tabular data (continued)}{tabular}

\oblock{keyword={tabular}, notkeyword={binary}, keyword={fca}, keyword={bmf}, keyword={categorical}, keyword={numerical}}{%
Transactional data can be seen as a binary matrix or table, and is hence a form of tabular data. Data tables might also contain categorical or real-valued attributes, which can be either turned into binary attributes as a pre-processing or handled directly with dedicated methods.
}

\rblock{More itemsets}{tabular-binary}{keyword={tabular}, keyword={binary}, notkeyword={tiles}}{%
  Beside \algKrimp{} and algorithms inspired from it, different approaches have been proposed to mine itemsets from binary matrices using the MDL principle.
  
\citet{heikinheimo_low-entropy_2009} focus on finding collections of low-entropy itemsets, typically even more compact than those obtained with \algKrimp{}.
On the other hand, the main objective of \citet{mampaey_summarising_2010} is to provide a summary of the dataset in the form of a partitioning of the items. For each partition, codewords are associated to the different combinations of the items that comprise the partition, and used to encode the corresponding subset of the data.
In the \algname{Pack} algorithm, proposed by \citet{tatti_finding_2008}, the code representing an item is made dependent on the presence/absence of other items in the same transaction. This can be represented as decision trees whose intermediate nodes are items and leaves contain code tables for other items.
\citet{fischer_discovering_2020} introduce a rich pattern language that can capture both co-occurrences and mutual exclusivity of items.

\citet{mampaey_summarizing_2012} present a variant of their \algname{MTV} itemset mining algorithm\noscite{~\citep{mampaey_tell_2011}}\seccite{ \xcfsec{beyond-maxent}} where the MDL principle is used to choose the collection of itemsets that yields the best model, as an alternative to the Bayesian Information Criterion (BIC).
}

\rblock{Blocks in binary data}{tabular-tiles}{keyword={tabular}, keyword={binary}, keyword={tiles}}{%
A different type of patterns that can be mined from binary matrices consists of blocks defined by a collection of rows and columns with a homogeneous density of ones, \doneedit{block patterns, overlap or no overlap, etc.}{sometimes known as \emph{tiles} or as \emph{biclusters}, }constituting the basis of \emph{block-based} strategies \xcfsec{basics-strategies}.
\citet{chakrabarti_fully_2004} propose a method to partition the rows and columns of the matrix into subsets that define homogeneous blocks. The example presented in \citfig{fig:blocks-binary} follows this approach. 
\citet{tatti_discovering_2012} introduce the \algname{Stijl} algorithm to mine a hierarchy of tiles from a binary matrix.
The adjectives \emph{geometric} or \emph{spatial} typically refer to cases where the order of the rows and columns of the data matrix is meaningful \citep[e.g.][]{papadimitriou_parameter-free_2005,faas_vouw_2020}.
}

\rblock{Formal Concept Analysis (FCA)}{tabular-fca}{keyword={tabular}, keyword={fca}}{%
The field of Formal Concept Analysis focuses on basically the same problem as itemset mining, but from a slightly different perspective and with its own formalism. Some steps have been taken towards applying and further developing MDL-based methods within this framework \citep{otaki_edit_2015,yurov_turning_2017}. In particular, as a topic in her doctoral dissertation, \citet{makhalova_contributions_2021} extensively studied MDL-based itemset mining methods from the perspective of FCA \citep[also][]{makhalova_first_2018,makhalova_mdl_2018,makhalova_coupling_2019,makhalova_likely-occurring_2021}. 
}

\rblock{Boolean Matrix Factorisation (BMF)}{tabular-bmf}{keyword={tabular}, keyword={bmf}}{%
  Factorising a matrix consists in identifying two matrices, the factors, so that the original matrix can be reconstructed as their product. One challenge is to find the right balance between the complexity of the decomposition (generally measured by the size of the factors) and the accuracy of the reconstruction, which is where the MDL principle comes in handy \citep{miettinen_model_2011,miettinen_mdl4bmf_2014,hess_primping_2017}. Whereas the other approaches permit reconstruction errors in both values, \citet{makhalova_from-below_2019} do not allow factors to cover zero entries, considering so-called ``from-below'' factorisations \citep[also][]{makhalova_from-below_2021}. \citet{lucchese_unifying_2014} \citep[also][]{lucchese_mining_2010,lucchese_generative_2010} propose a MDL-based score to compare factors of a fixed size.

An example of a factorisation of the toy binary dataset from \citfig{fig:data-binary-ex} is provided in \citfig{fig:binary-bmf}. The model consists of a pair of factor matrices whereas the data is encoded as a mask of corrections. To reconstruct the original data, the Boolean matrix product of the factors is computed and the mask of corrections is applied to the resulting matrix.
  
When applied to a Boolean matrix, factorisation shares some similarities with itemset mining, as it aims to identify items that occur (or do not occur) frequently together. That is, the two factor matrices can be interpreted as specifying itemsets and indicators of occurrence, respectively. On the other hand, the factors can also be interpreted as specifying possibly overlapping fully dense blocks. A mask applied to the reconstructed data provides a global error correction mechanism. Thus, Boolean matrix factorisation can be seen as a hybrid approach.
\qdone{added note about hybrid characteristics}
\qdone{added figure}

\begin{figure}[tbp]
\begin{tabular}{@{\hspace{0.15\textwidth}}c@{\hspace{0.05\textwidth}}c@{\hspace{0.05\textwidth}}c@{\hspace{0.15\textwidth}}}
\multicolumn{3}{l}{$L(M,D) =$} \\[-1em]
  $L(M)$ & $+$ & $L(D\mid{}M)$ \\
  \hspace{.35\textwidth} & & \hspace{.35\textwidth} \\[-.5em]
  \BMFExModel{} & \phantom{$+$} &  \BMFExData{} \\
\end{tabular}
\caption{Boolean Matrix Factorisation, example on the toy binary dataset of \citfig{fig:data-binary-ex}. The model $M$ consists of a pair of factor matrices (left) and the data $D$ is encoded as a mask of corrections (right). To reconstruct the original data, first the Boolean matrix product of the factors is computed, resulting in the matrix with two fully dense blocks $B_1$ and $B_2$. Then, the mask of corrections is applied to this matrix, that is, the value of the cells indicated in the mask (grey squares) are flipped. Note that the blocks drawn with the hatch pattern depict the intermediate step of the reconstruction and are not actually encoded as such.}
\label{fig:binary-bmf}
\end{figure}
}

\rblock{Categorical data}{tabular-categorical}{keyword={tabular}, keyword={categorical}}{%
One way to mine datasets involving categorical attributes is to binarise them and then apply, for instance, an itemset mining algorithm. However, binarisation entails a loss of information and dedicated methods can hence offer a better alternative.

\citet{mampaey_summarizing_2013} introduce an approach to detect correlated attributes, i.e.\ such that the different categories occur in a coordinated manner, \droppededit{more strictly speaking, looking for biclusters, possibly overlapping}{whereas \citet{he_relevant_2014} present a subspace clustering method}, i.e.\ look not only for groups of attributes, but also corresponding groups of rows where coordinated behaviour occurs.
}

\rblock{Numerical data}{tabular-numerical}{keyword={tabular}, keyword={numerical}}{%
A numerical data table containing $m$ columns can be seen as a collection of points in a $m$-dimensional space.
In pattern mining, numerical data is often handled by applying discretisation, which requires to partition the data into coherent blocks.
While it can be seen as a data exploration task in its own right, providing an overview of the dataset and the distribution of values, discretisation often constitutes a pre-processing task to allow applying algorithms that can handle only discrete input data.
Yet, choosing good parameters for the discretisation can be difficult, and its quality can have a major impact on later processing.
Unsupervised discretisation, where no side information is available, is in contrast to supervised discretisation, that takes into account class labels and often precedes a machine learning task. Here we focus on the former.

\citet{kontkanen_mdl_2007} propose a histogram density estimation method that relies on the MDL principle, formalised using the normalised maximum likelihood (NML) distribution. This method is employed by \citet{kameya_time_2011} to discretise time-series seen as a collection of two-dimensional time--measurement data points, and extended by \citet{yang_unsupervised_2020} to two-dimensional numerical data, more in general.
Along similar lines, \citet{nguyen_unsupervised_2014} aim to automatically identify a high-quality discretisation that preserves the interactions between attributes.
\citet{witteveen_realkrimp_2014} extend the Kraft inequality \xcfsec{basics-encoding} to numerical data and introduce an approach to find hyperintervals, i.e.\ multidimensional blocks.

\citet{makhalova_numerical_2019} consider the problem of mining interesting hyperrectangles from discretised numerical data, and aim to design an encoding that accommodates overlaps between patterns \citep{makhalova_mint_2020,makhalova_mint_2022}.

\citet{lakshmanan_generalized_2002} formalise mining OLAP data, i.e.\ multidimensional datasets, as a problem of finding a cover in a multidimensional array containing positive, negative and neutral cells. The aim is then to find the most compact set of hyperrectangles that covers all positive cells, none of the negative cells, and no more than a chosen number of the neutral cells. The score is presented as a generalised MDL due to the tolerance on neutral cells. However, coverings are evaluated by simply counting cells, which does not actually adhere with the principle, generalised or otherwise.
}
\end{srvsection}


\begin{srvsection}{Graphs}{graphs}
  
\oblock{keyword={graphs}, notkeyword={blocks}, notkeyword={hyperbolic}, notkeyword={substructures}, notkeyword={pathways}, notkeyword={infomap}}{%
In this section, we consider approaches for mining graphs. At their simplest, graphs are undirected and unlabelled, but they can also come with directed edges, with node or edge labels, or be dynamic, that is, time-evolving.
The main tasks consist in identifying nodes that have similar connection patterns to group them into homogeneous blocks and in finding recurrent connection substructures. These correspond respectively to \emph{block-based} and \emph{dictionary-based} strategies \xcfsec{basics-strategies}.

\qdone{added figure}
For illustrative purposes, we consider a toy graph, shown in \citfig{fig:data-graph-ex}, and delineate approaches that follow either strategy. The example shown in \citfig{fig:blocks-graph} illustrates the \emph{block-based} strategy and follows the work of \citet{chakrabarti_autopart_2004} \xcfsec{graphs-blocks}, whereas the example shown in \citfig{fig:dictionary-graph} illustrates the \emph{dictionary-based} strategy and follows the work of \citet{bariatti_graphmdl_2020} \xcfsec{graphs-substructures}.

\begin{pfigure}
  \begin{tabular}{@{}c@{\hspace{0.15\textwidth}}c@{}}
  \OrgGraphAdj{} & \OrgGraph{} \\
\end{tabular}
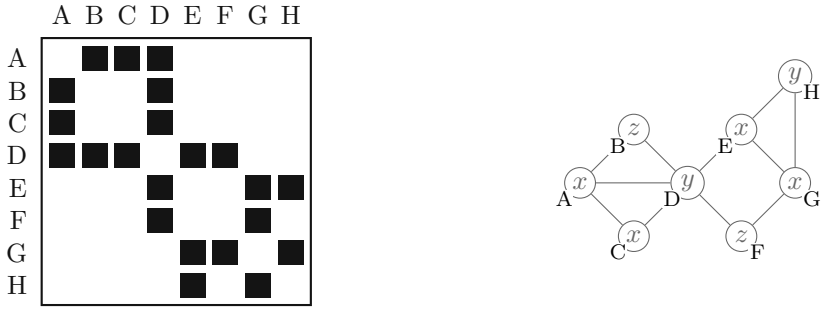
\caption{A toy graph, with eight nodes denoted $A$--$H$ carrying labels from $\{x, y, z\}$ and connected by undirected edges (right). The adjacency matrix (left) captures the structure of the graph, ignoring the labels.}
\label{fig:data-graph-ex}
\end{pfigure}

\begin{pfigure} \begin{tabular}{@{\hspace{0.15\textwidth}}c@{\hspace{0.02\textwidth}}c@{\hspace{0.02\textwidth}}c@{\hspace{0.15\textwidth}}}
\multicolumn{3}{l}{$L(M,D) =$} \\[-1em]
  $L(M)$ & $+$ & $L(D\mid{}M)$ \\
 \hspace{.3\textwidth} & & \hspace{.3\textwidth} \\[-.5em]
  \GBlocksExModel{} & \phantom{$+$} & \GBlocksExData{} \\
\end{tabular}
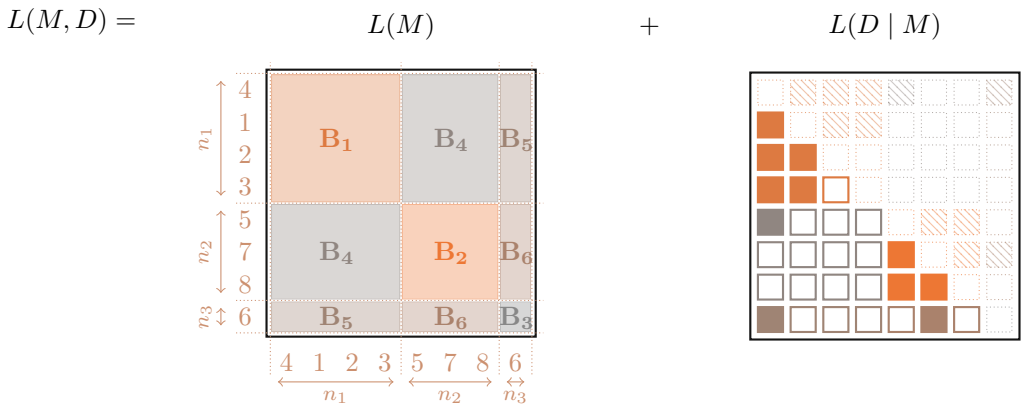
\caption{Block-based strategy, example on the toy graph of \citfig{fig:data-graph-ex}. Ignoring labels and considering its adjacency matrix, the graph can be encoded in a very similar way as a binary tabular dataset (see Section~\ref{sec:basics-strategies} and \citfig{fig:blocks-binary}). Furthermore, since the graph is undirected, and its matrix hence symmetric, it is enough to encode the lower triangular part of the matrix (depicted with solid lines and colour fill), from which the upper triangular part (depicted with dotted lines and hatch pattern) can be reconstructed. More intense shades of orange represent higher probabilities of ones within the corresponding block.}
\label{fig:blocks-graph}
\end{pfigure}

\begin{pfigure}
\begin{tabular}{@{}c@{}c@{}c@{}}
\multicolumn{3}{l}{$L(M,D) =$} \\[-1em]
  $L(M)$ & $+$ & $L(D\mid{}M)$ \\
  \hspace{.52\textwidth} & & \hspace{.42\textwidth} \\[-.5em]
  \GDictExModel{} & \phantom{$+$} &  \GDictExData{} \\
\end{tabular}
\caption{Dictionary-based strategy, example on the toy graph of \citfig{fig:data-graph-ex}.
The idea is similar to the one for binary tabular datasets (see Section~\ref{sec:basics-strategies} and \citfig{fig:dictionary-binary}). However, it is not enough to simply replace the subgraph patterns by their assigned codewords (depicted as blue blocks), the information about how the subgraphs are connected also needs to be encoded. Here, this is done through \emph{ports} associated to the patterns and their assigned codewords (depicted as tan blocks).}
\label{fig:dictionary-graph}
\end{pfigure}


Looking at the corresponding adjacency matrix, a simple unlabelled graph can be represented as a binary table. Approaches from Sections~\ref{sec:itemsets} and~\ref{sec:tabular} can thus readily be used for mining graphs.
On one hand, the problem of grouping nodes into blocks that constitute particularly dense subgraphs, or communities, is closely related to identifying particularly dense tiles in a binary matrix.
On the other hand, approaches that follow a \emph{dictionary-based} strategy and aim to identify substructures in the graphs share similarities with their counterparts for binary tabular data. However, it is not enough to simply replace the subgraph patterns by their assigned codewords. The information about how the subgraphs are connected also needs to be encoded, requiring more complex encoding schemes.

The survey by \citet{liu_graph_2018} covers graph summarisation methods in general, whereas \citet{feng_information-theoretic_2015} presents various information-theoretic graph mining methods, both of which include MDL-based methods for analysing graphs as a subset.
}

\rblock{Grouping nodes into blocks}{graphs-blocks}{keyword={graphs}, keyword={blocks}, notkeyword={dynamic}}{%
\citet{rosvall_information-theoretic_2007} present an information-theoretic framework to identify community structure in networks by grouping nodes and propose to use the MDL principle to automatically select the number of groups in which to arrange the nodes.

\citet{chakrabarti_autopart_2004} proposes to compress the adjacency matrix of a graph by grouping nodes into homogeneous blocks (see \citfig{fig:blocks-graph}), with a top-down procedure to search for a good partition. 
\citet{navlakha_graph_2008} similarly propose to build graph summaries by grouping nodes into supernodes, but with a bottom-up search procedure. 
A superedge linking two supernodes represents edges between all pairs of elementary nodes from either supernodes (hence a supernode with a loop represents a clique). When reconstructing the original graph, after expanding the supernodes and superedges, some corrections must be performed, to add and remove spurious edges.
\questionmrk{navlakha_graph_2008}{graphs-blocks}{extra piece}{Not sure why the sign of corrections is needed}%
\questionmrk{navlakha_graph_2008}{graphs-blocks}{unit cost}{Cost is number of encoded elements}%
\citet{navlakha_graph_2008} let the cost of encoding a graph equal the number of superedges and edge corrections, ignoring the cost of the assignment of nodes to supernodes.

\citet{khan_set-based_2015-1} \citep[also][]{khan_set-based_2015,khan_lossless_2015} work with essentially the same encoding, using locality-sensitive hashing (LSH) to identify candidates for merger, additionally considering node labels \citep{khan_set-based_2014}.
\questionmrk{khan_set-based_2015-1}{graphs-blocks}{missing piece}{Though it is unclear how they are incorporated into the encoding if at all. Big overlap between the articles.}%
\citet{akoglu_pics_2012} also aim at grouping nodes while taking into account node attributes.

A similar block summary approach for bipartite graphs is proposed by \citet{feng_summarization-based_2012}, with a more complete encoding.
\citet{papadimitriou_hierarchical_2008} also focus on bipartite graphs, but the obtained blocks are arranged into a hierarchy, while \citet{he_pack_2009} consider $k$-partite graphs.

In order to compress the adjacency matrix of an input graph more efficiently, \citet{he_automatically_2011} look for nodes with similar connection patterns, corresponding to similar rows in the matrix, and encode the differences, possibly in a recursive manner. The approach is used to spot nodes with unusual connections patterns, that do not lend themselves to grouping.

\citet{lefevre_grass_2010} propose a supernode summary involving superedge weights that represent the probability that an edge exists for each pair of nodes in the incident supernodes. In one variant of the problem, the MDL principle is used to choose the number $k$ of supernodes that strikes the best balance between model complexity ($k$) and fit to the data (reconstruction error).
\citet{lee_ssumm_2020} also consider summarising graphs by grouping nodes together, but fix a maximum length for the description of the model, i.e.\ the hypernodes, and look for the summary that minimises the reconstruction error, measured as the length of the description of edge corrections.

\citet{plant_data_2020} use a MDL-inspired score to learn graph embeddings. That is, the aim is to project the nodes into a multi-dimensional space, so that the structure of the graph is preserved as much as possible and, more specifically, such that connected nodes are placed close to each other. Therefore, the distance between any pair of nodes in the embedding is used to compute a probability that the nodes are connected, which, in turn is used to encode the presence or absence of the corresponding edge. Then, the quality of an embedding can be measured by how much it allows to compress the adjacency matrix.
}

\rblock{Grouping nodes into blocks in dynamic graphs}{graphs-blocks-dyn}{keyword={graphs}, keyword={blocks}, keyword={dynamic}}{%
\citet{sun_graphscope_2007} introduce a block summary approach for dynamic graphs, extended to multiple dimensions or contexts by \citet{jiang_catchtartan_2016}.
\citet{araujo_com2_2014} also propose a block summary approach for dynamic graphs, which they later extend to multiple dimensions or contexts as represented by qualitative labels on the edges \citep{araujo_discovery_2016}.
}

\rblock{Finding hyberbolic communities}{graphs-hyperbolic}{keyword={graphs}, keyword={hyperbolic}}{%
Instead of looking for blocks of uniform density and motivated by the observation that node degrees in real-world networks often follow a power-law distribution, \citet{araujo_beyond_2014} propose the model of \emph{hyperbolic communities}.
The name refers to the fact that when nodes in such communities are ordered by degree, edges in the adjacency matrix mostly end up below a hyperbola.

\citet{kang_beyond_2011} \citep[also][]{lim_slashburn_2014} decompose the input graph into hubs and spokes, with superhubs connecting the hubs recursively, and introduce a cost to evaluate how well the decomposition allows to compress the graph. This type of decomposition is proposed as an alternative to a decomposition into cliques, referred to as ``cavemen communities''.
}

\rblock{The Map Equation}{graphs-mapequation}{keyword={graphs}, keyword={infomap}}{%
\qdone{A section about the Map Equation suite of tools} 
\citet{rosvall_maps_2008} propose a method to reveal the important connectivity structure of weighted directed graphs. The approach assigns codes to nodes in such a way that random walks over the graph can be described succinctly. Furthermore, nodes are partitioned into modules so that the codes for nodes are unique within each module but can be reused between modules. A walk over the graph can then be described using a combination of codes indicating transitions between modules and lists of the successive nodes encountered within each module. The resulting two-level summary of the graph maps its main structures and the connections between and within them, and the approach is therefore referred to as the \emph{Map Equation}~\citep{rosvall_map_2009} or the \emph{Infomap} algorithm.\!\footnote{\url{https://www.mapequation.org/}}

Later, refinements and extensions of the method were proposed, to study changes in the connectivity structure over time~\citep{rosvall_mapping_2010}, reveal multi-level hierarchical connectivity structure~\citep{rosvall_multilevel_2011}, support overlaps between modules~\citep{viamontes_esquivel_compression_2011}, among others~\citep{bohlin_community_2014,de_domenico_identifying_2015,edler_mapping_2017,emmons_map_2019,calatayud_exploring_2019}. In particular, the method has found application in the analysis of ecological communities~\citep{edler_infomap_2017,blanco_punctuated_2021,rojas_multiscale_2021}.
}

\rblock{Identifying substructures}{graphs-substructures}{keyword={graphs}, keyword={substructures}, notkeyword={dynamic}}{%
  \citet{cook_substructure_1994} \citep[also][]{ketkar_subdue_2005} propose the \algname{Subdue} algorithm to mine substructures from graphs, possibly with labels, using the MDL principle. A substructure of the graph can be encoded and its occurrences in the graph be replaced by a single node. This can be done recursively, generating a hierarchical summary of the original graph. There are two shortcomings to the approach. First, replacing the substructure by a single node does not preserve the complete information about the connections to neighbours. Second, the matching of substructures is done in an approximate way, with an arbitrary fixed cost, rather than a proper encoding of the reconstruction errors (using the MDL principle for this evaluation is left for future work). Substructures are scored individually rather than in combination.
\questionmrk{cook_substructure_1994}{graphs-substructures}{other}{replacing the substructure by a single node does not preserve the complete information about the connections to neighbours. Second, the matching of substructures is done in an approximate way, with an arbitrary fixed cost, rather than a proper encoding of the reconstruction errors (using the MDL principle for this evaluation is left for future work)}%
\questionmrk{cook_substructure_1994}{graphs-substructures}{individual score}{Substructures are scored individually rather than in combination}%
The \algname{Subdue} algorithm is used by \citet{jonyer_mdl-based_2004} for the induction of context-free grammars, and by \citet{bloem_compression-based_2013} in comparative experiments against practical data compression with the \algname{GZIP} algorithm.

\citet{bloem_tutorial_2018} \citep[also][]{bloem_large-scale_2020} propose to use the MDL principle when evaluating the statistical significance of the presence of substructures in a graph.

The \algname{VoG} algorithm presented by \citet{koutra_vog_2014} \citep[also][]{koutra_summarizing_2015} allows to decompose the graph into basic primitives such as cliques, stars, and chains, which can overlap on nodes (but not on edges). Error corrections are then applied, to add and remove spurious edges. This can be seen as a global use of primitives. 
\citet{liu_empirical_2015} \citep[also][]{liu_reducing_2016} use the MDL principle to compare the ability of \algname{VoG} and graph clustering methods to generate graph summaries.
\citet{liu_reducing_2018} build on \algname{VoG} and address some of the shortcomings, such as the bias towards star structures, the inability to exploit edge overlaps, and the dependency on candidate order.
\citet{goebl_megs_2016} introduce a similar approach, with some of the same primitives, but prohibiting overlaps with the aim to make visualisation and interpretation easier.
The approach presented by \citet{bariatti_graphmdl_2020} removes the limitation to a predefined set of primitives and considers labelled 
graphs (see \citfig{fig:dictionary-graph}). The authors later upgraded the approach by generating candidates on-the-fly, thereby providing an anytime mining algorithm \citep{bariatti_graphmdl_2021}, and proposed a visualisation tool for the obtained graph patterns \citep{bariatti_graphmdlviz_2020}. This work forms the basis of a doctoral dissertation \citep{bariatti_mining_2021}. 

The approach proposed by \citet{coupette_graph_2021} aims to highlight similarities and differences between graphs, and is akin in spirit to the \algname{DiffNorm} algorithm for transactional data \xcfsec{itemset-comp}. It looks for a common model consisting of basic primitives, like those used in \algname{VoG}, such that each graph can be reconstructed based on these primitives adjusted through parameters specific to the graph, as well as additional structures, where necessary.
\qdone{\citet{coupette_graph_2021} diffnorm for graphs}

\citet{feng_compression-based_2013} exploit basic structures of graphs, like stars and triangles, to save on the encoding of the adjacency matrix. Such primitives assign a probability to the existence of an edge, which is used to encode it. Which primitive applies is determined in part based on structure information available so far, i.e.\ previously decoded. This can be seen as a local use of primitives.

\citet{belth_what_2020} learn relational rules which can be used to summarise knowledge graphs, involving typed edges and nodes.
}

\rblock{Identifying substructures in dynamic graphs}{graphs-substructures-dyn}{keyword={graphs}, keyword={substructures}, keyword={dynamic}}{%
\citet{shah_timecrunch_2015} \citep[also][]{shah_summarizing_2017} extend the \algname{VoG} approach to dynamic graphs. More specifically, they incorporate the temporal aspect of substructures appearing only at given time steps, across a range of contiguous time steps, periodically, or in a flickering fashion. Therefore, in addition to decomposing the graph into basic structures, one needs to indicate when these structures appear. 
The \algname{Mango} algorithm by \citet{saran_summarizing_2019} also looks for predefined structures in a dynamic graph, aiming more specifically at tracking their evolution through time.
}

\rblock{Finding pathways between nodes}{graphs-pathways}{keyword={graphs}, keyword={pathways}}{%
Given a large graph, \citet{akoglu_mining_2013} consider the problem of identifying a set of marked nodes. This can be done by listing the node identifiers or by navigating between nodes. The latter strategy requires to choose between the limited number of neighbours of each traversed node, rather than among all possible nodes in the graph, potentially leading to shorter descriptions. In particular, the problem formulated by \citet{akoglu_mining_2013} consists in finding the best collection of trees spanning the marked nodes in the graph. The graph as such is not encoded, it is regarded as shared knowledge.

\citet{prakash_efficiently_2014} similarly assume shared knowledge of the graph. The aim is then to transmit the starting points and spread of an epidemic through the graph over a sequence of time steps, assuming a ``susceptible--infected'' (SI) epidemic model.
}
\end{srvsection}


\begin{srvsection}{Temporal data}{sequences}
  
\oblock{keyword={sequential}, notkeyword={segmentation}, notkeyword={substring}, notkeyword={subsequences}, notkeyword={motifs}, notkeyword={periodicity}, notkeyword={trajectories}}{%
In this section, we look at data where the attribute values come as a sequence, i.e.\ in a specific order. In particular, this order might correspond to time, in which case the data is called \emph{temporal}. 
In some cases only the order matters, whereas in other cases absolute positions are associated to the values, such as timestamps in temporal data.
In addition to time, spatial dimension(s) might be associated to the values, resulting in \emph{spatio-temporal} data.
The terms \emph{sequential data} and \emph{sequence} are sometimes used to refer more narrowly to sequences of discrete attributes or items, which are typically called \emph{events}. 
On the other hand, the term \emph{timeseries} is generally used to refer to real-valued attributes sampled at regular or irregular time intervals.
Text and genetic data (such as DNA or RNA sequences) fall into the former category. More specifically, such data generally comes in the form of \emph{strings}, that is, as sequences of characters that represent occurrences of single items where the order is meaningful, not the positions.
The data might consist of a single long sequence or of a database of multiple, typically shorter, sequences.

As with other types of data, most of the work on mining sequential data can be divided into two main tasks, namely segmentation and frequent pattern mining, corresponding to \emph{block-based} and \emph{dictionary-based} strategies, respectively \xcfsec{basics-strategies}.
In segmentation problems (a.k.a.\ change point detection), the aim is to divide the input data into homogeneous blocks or segments, each associated to specific occurrence probabilities of the different events. 
On the other hand, in frequent pattern mining, the aim is to find recurrent substructures, which are commonly referred to as \emph{episodes} and \emph{motifs} when considering sequences and timeseries, respectively.

\qdone{added figure}
For illustrative purposes, we consider a toy sequence, shown in \citfig{fig:data-seq-ex}, and delineate approaches that follow either strategy. The example shown in \citfig{fig:blocks-seq} illustrates the \emph{block-based} strategy and follows the work of \citet{kiernan_constructing_2008} \xcfsec{sequential-seg-sequences}, whereas the example shown in \citfig{fig:dictionary-seq} illustrates the \emph{dictionary-based} strategy and follows the work of \citet{tatti_long_2012} \xcfsec{sequential-subsequence}.
While similar to their counterparts for binary tabular data, approaches for temporal data must account for the order that the special dimension of time imposes on the occurrences.

\begin{pfigure}
  \layoutnarw{\hspace*{.7em}}\OrgSeq{}
  
  \layoutnarw{\hspace*{.7em}}\OrgSeqMat{}
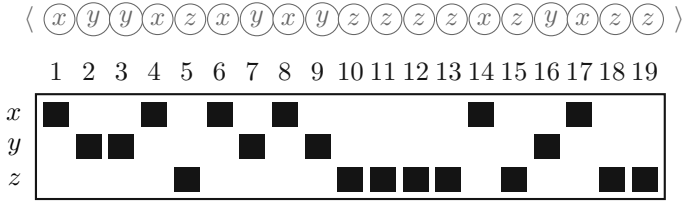
\caption{A toy sequence, with nineteen consecutive occurrences over three distinct events \evtA{}, \evtB{} and \evtC{} (top), also represented as a binary matrix (bottom) indicating which event (row) occurs at a given time step (column).}
\label{fig:data-seq-ex}
\end{pfigure}

\begin{pfigure}

\begin{tabular}{@{}c@{}c@{}c@{}}
\multicolumn{3}{l}{$L(M,D) =$} \\[-1em]
  $L(M)$ & $+$ & $L(D\mid{}M)$ \\
\layoutnarw{%
\hspace{.54\textwidth} & & \hspace{.45\textwidth} \\[-.5em]
  \multicolumn{3}{l}{\SBlocksExModel{}} \\ 
  \multicolumn{3}{r}{\SBlocksExData{0}} \\
}%
\layoutwide{%
  \\[-.5em]
  \SBlocksExModel{} & \multicolumn{2}{r}{\SBlocksExData{\guideybSBlocksData}} \\
}%
\end{tabular}
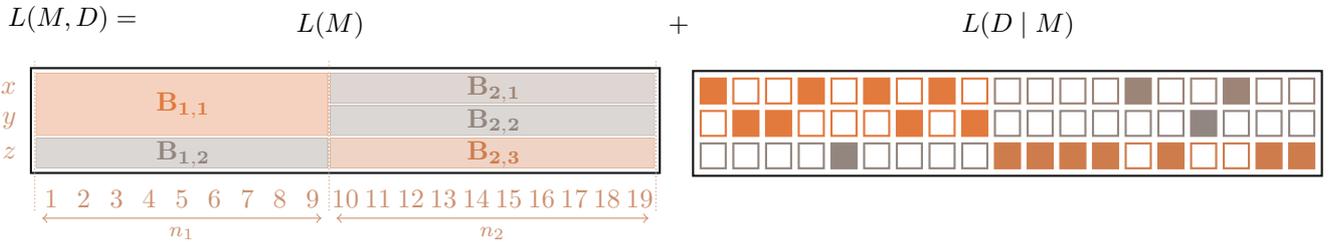
\caption{Block-based strategy, example on the toy sequence of \citfig{fig:data-seq-ex}. The sequence can be encoded in a very similar way as a binary tabular dataset (see Section~\ref{sec:basics-strategies} and \citfig{fig:blocks-binary}). In this case, however, the order of the columns corresponds to time and is therefore fixed, but the events (i.e.\ rows) can be arranged into different groups in the different time segments (i.e.\ column groups). More intense shades of orange represent higher probabilities of ones within the corresponding block.}
\label{fig:blocks-seq}
\end{pfigure}

\begin{pfigure}
\begin{tabular}{@{}c@{}c@{}c@{}}
\multicolumn{3}{l}{$L(M,D) =$} \\[-1em]
  $L(M)$ & $+$ & $L(D\mid{}M)$ \\
\layoutnarw{%
  \hspace{.55\textwidth} & & \hspace{.45\textwidth} \\[-.5em]
  \multicolumn{1}{r}{\SDictExModel{5}} & \\
  \multicolumn{3}{r}{\SDictExData{0}{2}{14}{2}} \\
}%
\layoutwide{%
  \hspace{.45\textwidth} & & \\[-.5em]
  \SDictExModel{\lgdxSDictModel} &
  \multicolumn{2}{r}{\SDictExData{\guideybSDictData}{\guideytSDictData}{\lgdxSDictData}{\lgdySDictData}} \\
}
\end{tabular}
\caption{Dictionary-based strategy, example on the toy sequence of \citfig{fig:data-seq-ex}.
The idea is similar to the one for binary tabular datasets (see Section~\ref{sec:basics-strategies} and \citfig{fig:dictionary-binary}). Episode patterns are assigned codewords (depicted as blue blocks). The original sequence is reconstructed by reading codewords from the \emph{pattern stream} ($\mathcal{P}$) and inserting the events of the corresponding episode into the sequence, in order. In addition, each multi-event episode in the code table is associated to an additional pair of codewords (depicted as bronze blocks). These codewords, read from the \emph{gap stream} ($\mathcal{G}$), indicate whether to insert the next element from the current episode (\evtFill) or to leave a gap where to insert the next episode (\evtGap).}
\label{fig:dictionary-seq}
\end{pfigure}


}

\rblock{Finding haplotype blocks}{sequential-seg-haplotype}{keyword={sequential}, keyword={segmentation}, keyword={haplotype}}{%
A haplotype, or haploid genotype, is a group of alleles of different genes on a single chromosome, which are closely linked and typically inherited as a unit. 
Several works have been dedicated to the problem of finding haplotype block boundaries, i.e.\ identifying block structure in genetic sequences\noscite{ \citep{koivisto_mdl_2002,anderson_finding_2003,greenspan_model-based_2003,mannila_minimum_2003,greenspan_model-based_2004}}. This requires jointly partitioning multiple aligned strings.
}

\rblock{Segmenting sequences}{sequential-seg-sequences}{keyword={sequential}, keyword={segmentation}, keyword={sequences}}{%
Several approaches have also been developed for segmenting event sequences more in general.

The method introduced by \citet{kiernan_constructing_2008} partitions a sequence into time segments, then partitions the events of each segment into groups (see \citfig{fig:blocks-seq}). The proposed algorithm is then extended to allow overlaps and gaps between segments \citep{kiernan_constructing_2009} and a tool to visualise the obtained segmentation is proposed \citep{kiernan_eventsummarizer_2009}.
\citet{wang_algorithmic_2010} further aim to model dependencies between segments.

The algorithm proposed by \citet{lam_decomposing_2014} partitions the alphabet into subsets, then separately encodes the sequence projected on each subset of symbols, as well as a sequence that maps each position to the corresponding subset.
\questionmrk{lam_decomposing_2014}{sequential-seg-sequences}{missing piece}{The encoding does not specify which symbol belongs to which subset, so there is some loss}%
\questionmrk{lam_decomposing_2014}{sequential-seg-sequences}{which part}{the term that encodes the assignment of positions to subsets is presented as the cost of the model, but it can be seen as part of encoding the data given the model, the model is the assignment of symbols to subsets.}%
\citet{chen_automatic_2018} adopt a generic point of view on sequence segmentation, considering that the input data can be either univariate or multivariate, consist of categorical or real-valued variables, with no assumption on the underlying distribution.
\citet{gautrais_widening_2020} aim to segment a sequence in such a way that each segment contains a collection of recurrent ``signature'' events. In particular, they consider retail data and apply the approach to the sequences of transactions of individual customers, in order to analyse their shopping behaviour. 
}

\rblock{Segmenting timeseries}{sequential-seg-timeseries}{keyword={sequential}, keyword={segmentation}, keyword={timeseries}}{%
\citet{hu_discovering_2011} \citep[also][]{hu_towards_2013,hu_using_2015} propose an approach to represent a timeseries with an Adaptive Piecewise Constant Approximation (APCA) and define the intrinsic cardinality of the data to be the number of distinct constant values in the approximation. In this approach, a timeseries is encoded by specifying the end-points and value of each segments, then listing reconstruction errors.
\questionmrk{hu_discovering_2011}{sequential-seg-timeseries}{missing piece}{This does not account for the transmission of errors code table.}%
\questionmrk{hu_discovering_2011}{sequential-seg-timeseries}{code choice}{Besides, Huffman coding rewards having few different recurrent error values, not necessarily small ones.}%
\citet{vespier_mdl-based_2012} also decompose timeseries into segments, but consider components at multiple time-scales, modeled as piecewise constant or polynomial functions.

\citet{matsubara_autoplait_2014} consider the problem of segmenting multivariate timeseries. A timeseries is modeled using a multi-level hidden Markov model (HMM), where high-level states represent regimes that contain lower level states.
\questionmrk{matsubara_autoplait_2014}{sequential-seg-timeseries}{missing piece}{How this approach handles and reconstructs real values is unclear.}%
\citet{wu_modeling_2020} also consider a problem of timeseries segmentation, and model each segment with a Markov chain.
\questionmrk{wu_modeling_2020}{sequential-seg-timeseries}{other}{Authors say the model is encoded lossily, but not clear where the loss happens.}%

\citet{rakthanmanon_time_2011} \citep[also][]{rakthanmanon_mdl-based_2012} consider the problem of clustering sequential data, in particular such as arises from discretised timeseries where each distinct numerical value is mapped to a distinct symbol. The authors argue that not every value is of interest, because sequences tend to contain meaningless transitions, and that the MDL principle can help identify the segments of interest.
\citet{begum_towards_2013} \citep[also][]{begum_minimum_2014} use the clusters obtained with this approach for the task of semi-supervised classification.
}

\rblock{Mining substrings}{sequential-substring}{keyword={sequential}, keyword={substring}}{%
The algorithm devised by \citet{evans_improved_2006} \citep[also][]{markham_implementation_2009} searches for the best set of substrings to encode an input string according to the proposed Optimal Symbol Compression Ratio (OSCR) \citep{evans_new_2003}. The algorithm, which has been applied primarily to analyse genetic sequences \citep{evans_microrna_2007}, is iterative, at each step picking the substring that compresses most and replacing it by a temporary code. Selected substrings can be recursive, in the sense that they contain previously selected substrings. In the end, the selected substrings are assigned codes using Huffman coding. 
}

\rblock{Mining episodes from sequences}{sequential-subsequence}{keyword={sequential}, keyword={subsequences}}{%
\citet{lam_mining_2012} propose to encode timestamped sequences with absolute positioning. That is, the positions of covered occurrences are listed separately for each singleton event or selected subsequence. A fixed-length code is used, so all elements (event or position) cost the same and, in particular, occurrences can appear arbitrarily far apart with no penalty. 
\questionmrk{lam_mining_2012}{sequential-subsequence}{unit cost}{reportedly to avoid the issue of bit representation.}%
Follow-up work \citep{lam_mining_2014} focuses on strings. The proposed algorithms have the same names (\algname{SeqKrimp} and \algname{GoKrimp}), but use a different encoding mechanism. Specifically, having constructed a dictionary mapping subsequences to codewords, each match of a selected subsequence is replaced by its associated codeword, followed by Elias codes indicating the gaps between occurrences of the successive events of the subsequence. For a given subsequence, a subroutine is proposed to find the matches with minimum gap cost.
\questionmrk{lam_mining_2014}{sequential-subsequence}{missing piece}{When decoding the dictionary that associate each singleton event and selected subsequence to its codeword, it is not clear how the receiver knows where the codewords end, which is necessary to be able to reconstruct them.}%
\citet{lam_zips_2013} consider a similar problem in a streaming setting. The proposed encoding points back to the previous occurrence of the subsequence, with a flag to indicate when an extended subsequence should be recorded as new, that is, added to the dictionary.
\questionmrk{lam_zips_2013}{sequential-subsequence}{extra piece}{It is not clear why a dictionary is needed at all, why do patterns need to be associated to codewords? (explanation about dictionary D in section 3.2 and example 2 are somewhat contradictory. It seems the dictionary is needed to recover the patterns, not actually for encoding. This is not really two-parts MDL, since no model is actually being encoded).}%
\questionmrk{lam_zips_2013}{sequential-subsequence}{extra piece}{Reference is to the first event of pattern, why not the last one? It is closer, would it not allow interleaving the same pattern? (I think it works).}%
\questionmrk{lam_zips_2013}{sequential-subsequence}{extra piece}{Why point back to position of last occurrence, rather than keep track of order of last encountered patterns and refer to position on that list (i.e.\ repetition of the same pattern do not count multiple times and clutter)}%

The \algname{Sqs} (``squeeze'') algorithm of \citet{tatti_long_2012} follows a dictionary-based strategy and is similar to \algKrimp{} but for sequences (see \citfig{fig:dictionary-seq}). Each selected subsequence, or episode, is assigned a codeword representing it, as well as a pair of codewords representing gap (move to next position) and fill (insert event) operations. Gaps are allowed but not interleaving. In other words, gaps must be filled by singletons.
\questionmrk{tatti_long_2012}{sequential-subsequence}{other}{would we run the risk to confuse gap/fill codes of different patterns?}%
This work is then extended in multiple ways, to take into account an ontology over the events, resulting in algorithm \algname{Nemo} \citep{grosse_summarising_2017} and by adding support for rich interleaving and choice of events in patterns, resulting in algorithm \algname{Squish} \citep{bhattacharyya_efficiently_2017}.

After focusing on the analysis of seismic data, aiming to cluster and compare seismograms represented as multiple aligned sequences \citep{bertens_characterising_2014}, \citet{bertens_keeping_2016} consider multivariate event sequences more in general and propose algorithm \algname{Ditto}, which can be seen as an extension of \algname{Sqs} to handle multivariate patterns. The work constitutes the basis of a dissertation focused on detecting anomalies and mining multivariate event sequences, also in combination, i.e.\ employing \algname{Ditto} to detect anomalies in such sequences \citep[Chapter 7]{bertens_insight_2017}.
\citet{hinrichs_characterising_2017} use compression and the \algname{Sqs} algorithm to analyse the similarities between sequence databases in terms of occurring sequential patterns, focusing mostly on text data. The proposed algorithm, called \algname{SqsNorm}, provides for sequential data the type of analysis that \algname{DiffNorm} allows for transactional data \xcfsec{itemset-comp}.

The approach proposed by \citet{wiegand_mining_2021} is clearly related to \algname{Sqs} and \algname{Squish} but aims to summarise entire complex event sequences, rather than capturing fragmentary behaviour. The models considered resemble Petri nets or finite state machines and specify conditional transitions between events. The data is then represented as a succession of instructions that, when fed through the model, allow to reconstruct the original event log. The authors later present a similar model called event-flow graph \citep{wiegand_mining_2022}. Instead of pattern nodes, this model involves rules defined over attribute vectors associated to the sequences.
\qdone{Added \citet{wiegand_mining_2021}}

\citet{cuppers_just_2020} aim to identify sequential patterns that reliably predict the impending occurrence of an event of interest. In other words, they look for a set of rules, but in sequential rather than transactional data \xcfsec{itemset-rules}. Along similar lines, \citet{bourrand_discovering_2021} \citep[also][]{bourrand_discovering_2021-1} aim to discover a compact set of sequential rules from a single long event sequence.
\qdone{back ref to rules section}

\citet{fowkes_subsequence_2016} propose a generative probabilistic model of sequence databases. The authors discuss the connection between probabilistic modeling and description length, and compare their proposed algorithm, which is not based on the MDL principle, to \algname{Sqs} and \algname{GoKrimp}.

\citet{ibrahim_discovering_2016} consider sequences with timestamps and patterns that consist of subsequences with fixed inter-event times. The data is encoded by listing the patterns, along with their occurrences. More specifically, for each subsequence, the events and inter-event times are specified, as well as the timestamp of the first event of each occurrence.
\questionmrk{ibrahim_discovering_2016}{sequential-subsequence}{unit cost}{Everything is encoded as a list of integers, using a fixed-length code (to avoid penalising late occurrences).}%
\questionmrk{ibrahim_discovering_2016}{sequential-subsequence}{other}{Suggests to use Hamming code, to reduce encoding size. Resorting to error correcting code does not make sense in this context}%
\citet{mitra_summarizing_2019} propose a generative statistical model (a hidden Markov model, HMM) as justification for this encoding. 
\citet{yan_swift_2018} look for patterns in a stream of sequential data where each event is associated to a timestamp, using a sliding window and maximum gap constraint.
\questionmrk{yan_swift_2018}{sequential-subsequence}{unit cost}{The cost of a pattern is first defined to be the number of characters used to represent it, that is, the number of events plus the number of timestamps. In a second version of the approach, this cost is then defined to be equal to one for all patterns, reportedly to avoid bias against patterns involving more events.}%
}

\rblock{Mining motifs from timeseries}{sequential-motifs}{keyword={sequential}, keyword={motifs}}{%
  \citet{tanaka_discover_2003} \citep[also][]{tanaka_discovery_2005} look for motifs in timeseries, representing discretised values as integers using a fixed-length code.
\questionmrk{tanaka_discover_2003}{sequential-motifs}{missing piece}{however the codeword length is not known to receiver, who is thus not able to reconstruct the data even up to discretised values}%
  
\citet{shokoohi-yekta_discovery_2015} consider the problem of extracting rules from timeseries, aiming to match shapes rather than the precise values. The proposed score is used to evaluate the consequent of candidate rules, allowing to compare consequents of different lengths. The score evaluates the compression gain resulting from specifying the motif once and then listing the errors for each occurrence, instead of listing the actual values for each occurrence. This is applied to evaluate candidates individually, not as a set.
\questionmrk{shokoohi-yekta_discovery_2015}{sequential-motifs}{other}{Disclaimer, inspired by MDL, not purist.}%
}

\rblock{Mining periodic patterns}{sequential-periodicity}{keyword={sequential}, keyword={periodicity}}{%
Exploiting regularities not only about what happens, that is, finding coordinated event occurrences, but also about when it happens, that is, finding consistent inter-occurrence time intervals, can allow to further compress the data.

In the context of ``smart homes'' and health monitoring, \citet{heierman_mining_2004} look for periodically repeating events or sets of events in a sequence, with a MDL criterion to identify interesting candidates, which are then used to automatically construct a Markov model (HPOMDP) \citep[also][]{heierman_improving_2003,das_health_2004,youngblood_automated_2005}. 
\questionmrk{heierman_mining_2004}{sequential-periodicity}{missing piece}{No details about the encoding, only intuitions.}%
\questionmrk{heierman_mining_2004}{sequential-periodicity}{individual score}{Apparently, compression is used to compute a score for each candidate pattern individually.}%
The work of \citet{rashidi_com_2013} shares the same context and goal, further accounting for discontinuities in the repetitions and variations in the order of the events.
\questionmrk{rashidi_com_2013}{sequential-periodicity}{missing piece}{Also no details about encoding. Includes a continuity factor and a term to account for variations in the patterns, not sure how they are computed and what the equation means in terms of encoding.}%
\questionmrk{rashidi_com_2013}{sequential-periodicity}{individual score}{Code lengths are computed for each pattern separately and rescaled to the unit interval.}%

\citet{galbrun_mining_2018} introduce patterns involving nested periodic recurrences of different events and a method for constructing them by combining simple cycles into increasingly complex and expressive patterns.
}

\rblock{Trajectories}{sequential-trajectories}{keyword={trajectories}}{%
  \citet{phan_mining_2013} consider data that represent a collection of moving objects, each associated at each timestamp with a geospatial position. As a pre-processing step, the objects must be clustered based on proximity, separately for each timestamp. An object is allowed to belong to several clusters at any given timestamp. The goal is then to find a sequence of clusters (at most one per timestamp) having objects in common. The result is called a \emph{swarm} and intended to represent objects moving together.
\questionmrk{phan_mining_2013}{sequential-trajectories}{unit cost}{The proposed encoding uses a fixed-length code, i.e.\ scores by number of elements.}%
When encoding the trajectory of an object (as a sequence of clusters) patterns can be used whole, or from/to an intermediate position.
\questionmrk{phan_mining_2013}{sequential-trajectories}{missing piece}{It is not clear how objects that belong to multiple clusters for any given timestamp are handled.}%
  
Also considering spatio-temporal data but in a different scenario, \citet{zhao_clean_2019} aim to mine frequent patterns in trajectories over a road network. Mapping the trajectories to the corresponding road segments effectively turns the problem into a frequent sequence mining problem. The MDL principle is used to formulate a problem of trajectory spatial compression, addressed using a dictionary-based strategy.
\questionmrk{zhao_clean_2019}{sequential-trajectories}{unit cost}{fixed cost of one per element.}%
}
\end{srvsection}


\section{Discussion}
\label{sec:discussion}


Our goal here is to open a discussion on issues relevant to MDL-based methods for pattern mining.
In particular, the design of the encoding is a crucial ingredient when developing such a method. We consider different questions that might be raised by the involved choices, regarding, in particular, conformity and suitability with respect to the MDL principle. To illustrate the discussion, we point to various works listed in the previous sections.

\subsection{Encoding in question}
\label{sec:discuss-encoding}
Alleged infractions to the MDL principle can be of different kinds and degrees of severity.
They include cases where
\textit{(i)} the assignment of codewords ignores information theory,
\textit{(ii)} the proposed encoding is not functional due to some information missing, and
\textit{(iii)} the proposed encoding clearly cannot achieve a good compression due to the presence of unnecessary unjustified terms.

\medskip
Several methods assign the same, typically unit, cost to all encoded elements (which might be items, nodes, edges, events, timestamps, etc.\ depending on the case, or even entire patterns), so that the description length is simply the number of encoded elements (see for instance \cite{navlakha_graph_2008} \xinsec{graphs-blocks}, and \cite{phan_mining_2013,zhao_clean_2019} \xinsec{sequential-trajectories}). Some authors motivate this choice by the need to avoid penalising large values, or to circumvent other encoding issues (see for instance \cite{ibrahim_discovering_2016,lam_mining_2012} \xinsec{sequential-subsequence}).
In the method proposed by \citet{yan_swift_2018} \xcfsec{sequential-subsequence}, for instance, the cost of a pattern is first defined to be the number of characters used to represent it, that is, the number of events plus the number of timestamps. In a second version of the method, this cost is then defined to be equal to one for all patterns, reportedly to avoid bias against patterns involving more events.

Transmitting the dataset through a binary communication channel as efficiently as possible is a thought experiment of sorts that motivates the score.
If short codewords are assigned to specific elements because they are deemed more valuable and useful, then other elements will have to be assigned longer codewords, because not everything can be transmitted cheaply. One can think of it as the fundamental limits of information theory, through this compression scenario, forcing the designer of the method to make choices as to what he considers important and interesting. 
One might argue that using unit costs corresponds to using a fixed-length code and rescaling everything for convenience. This indeed simplifies the design of the encoding, as it avoids making decisions, and in this sense short-circuits the principle.

\medskip
Small coding elements, such as for example required to delimit patterns in the code table, are often omitted. This is sometimes done deliberately, putting forth, in particular, the use of a pre-defined framework to be filled with the relevant values, which is common to all models and can therefore be ignored (see for instance \cite{vreeken_krimp_2011} \xinsec{itemsets}, and \cite{van_leeuwen_association_2015} \xinsec{itemset-rules}), but is sometimes left unexplained and might seem accidental. In the approach proposed by \citet{lam_mining_2014} \xcfsec{sequential-subsequence}, it is unclear how the receiver knows where the codewords end when decoding the dictionary. On the other hand, \citet{tanaka_discover_2003} \xcfsec{sequential-motifs} use a fixed-length code to encode values from a set, but the number of distinct values, which varies for different models, is not transmitted, so that the receiver cannot deduce the codeword length, and hence cannot decode the message.

More substantial pieces might also be missing. For instance, the encodings proposed by \citet{lam_decomposing_2014} \xcfsec{sequential-seg-sequences} and by \citet{hu_discovering_2011} \xcfsec{sequential-seg-timeseries} do not account for the transmission of the assignment of symbols to subsets and of the mapping of offset values to codewords for the corrections, respectively, which are needed to reconstruct the data.

Explanations about the encoding are sometimes kept at the level of intuitions, and the details provided can be insufficient to properly understand how it works (see for instance \cite{khan_set-based_2015-1} \xinsec{graphs-blocks}, \cite{matsubara_autoplait_2014} \xinsec{sequential-seg-timeseries}, \cite{heierman_mining_2004,rashidi_com_2013} \xinsec{sequential-periodicity}, and \cite{phan_mining_2013} \xinsec{sequential-trajectories}).

Arguably, ensuring decodability would in some cases require only minor modifications of the encoding scheme, and would likely have no major impact on the results.
Furthermore, how much effort should be spent ensuring that the proposed encoding works is debatable, since it will never be used in practice.

\medskip
The choice of encoding can sometimes seem sub-optimal, ill-suited or introduce undesirable bias.
For instance, \citet{navlakha_graph_2008} \xcfsec{graphs-blocks} list edge corrections that should be applied to the reconstructed graph, indicating for each one the sign of the correction. It seems, however that this information is unnecessary, as it can be inferred from the reconstructed graph, by checking whether the edge is present (must be deleted) or absent (must be added).
\citet{hu_discovering_2011} \xcfsec{sequential-seg-timeseries} encode a list of value corrections using Huffman coding, meaning that having few distinct but recurrent error values is rewarded, not necessarily small ones. Using a universal code for the corrections would instead encourage small error values, which might be more intuitive. In any case, it is advisable to lay bare and motivate the potential biases introduced by the choice of encoding, whenever possible. 

Considering a sequence, \citet{lam_zips_2013} \xcfsec{sequential-subsequence} encode the occurrence of an event or subsequence by pointing back to the position of the first occurrence. Pointing back, instead, to the position of the last encountered occurrence would require to encode smaller values and might lead to savings. Keeping track of the order in which the patterns were last encountered and referring to the position in that list, so that repetitions of the same pattern do not fill up the list, is another alternative. There are often different ways to achieve the same purpose, not necessarily with a clear overall best choice.  
In addition to pointing back to previous occurrences, \citet{lam_zips_2013} maintain a dictionary of patterns. It is unclear whether the dictionary is actually needed for the encoding, or is primarily used to recover the encountered patterns.

What is part of the encoding of the model and what is part of the encoding of the data given the model is sometimes not entirely obvious. 
For example, the algorithm of \citet{lam_decomposing_2014} \xcfsec{sequential-seg-sequences} encodes a sequence by partitioning the alphabet and considering separately the subsequences over each subset of symbols. The authors present the term that corresponds to the assignment of positions to subsets as part of the encoding of the model. Debatably, it can be considered instead as part of the encoding of the data given the model, while the assignment of symbols to subsets, which is ignored, would belong to the encoding of the model. 
Besides, encodings often actually consist of three terms, \textit{(i)} a description of the set of patterns (the model), \textit{(ii)} information to reconstruct the data using these patterns, and \textit{(iii)} a list of corrections to apply to the reconstructed data in order to recover the original data, with the latter two together representing the data given the model.

\subsection{Code of choice}
\label{sec:discuss-code}
\qdone{Discussing the superiority prequential coding and one-part codes in general}
Prequential plug-in codes, and refined codes more in general, provide means to avoid unwanted bias arising from arbitrary choices in the encoding \xcfsec{basics}.

For instance, \citet{budhathoki_difference_2015} use prequential coding for the itemset occurrences in the \algname{DiffNorm} algorithm \xcfsec{itemset-comp}. The choice is especially relevant in this scenario where the goal is to contrast the itemset make-up of different datasets, and not to inspect the usage of itemsets in a particular dataset.
\citet{bhattacharyya_efficiently_2017} as well as \citet{wiegand_mining_2021} use prequential coding for the streams that contain information about pattern occurrences \xcfsec{sequential-subsequence}.
Other recent works \citep[][cf.\ Sections~\ref{sec:tabular-tiles}, \ref{sec:tabular-numerical} and \ref{sec:graphs-substructures}, respectively]{faas_vouw_2020,makhalova_mint_2020,bloem_large-scale_2020} also use prequential coding, while \citet{bertens_keeping_2016,hinrichs_characterising_2017} \xcfsec{sequential-subsequence} both explicitly suggest upgrading the current encoding with a prequential code, as a direction for future work.
Going further, \citet{proenca_robust_2021} improved on their earlier work \citep{proenca_interpretable_2020} \xcfsec{itemset-rules} by replacing prequential coding, which is only asymptotically optimal, with normalised maximum likelihood (NML), which is optimal for fixed sample sizes, employing similar techniques as \citet{kontkanen_mdl_2007} \xcfsec{tabular-numerical}.

However, modern Bayesian and NML codes can be challenging to compute, or even downright infeasible. Furthermore, one-part codes can be less intuitive than two-part codes, and do not provide as direct an access to information about pattern usage.
For instance, \citet{mampaey_summarising_2010} \xcfsec{tabular-binary} compare two encodings, with and without prequential coding, and, obtaining similar results, choose to proceed with the latter as it is more intuitive.
All in all, modern refined codes have improved theoretical properties, but using them to build better methods comes with some challenges.

\subsection{The letter or the spirit}
\label{sec:discuss-spirit}
Some approaches use the MDL principle to score and compare individual candidate patterns, rather than evaluating them in combination (see for instance \cite{cook_substructure_1994} \xinsec{graphs-substructures}, \cite{shokoohi-yekta_discovery_2015} \xinsec{sequential-motifs}, as well as \cite{heierman_mining_2004,rashidi_com_2013} \xinsec{sequential-periodicity}).

Considering a two-view dataset, i.e.\ a dataset consisting of two tables, the approach proposed by \citet{van_leeuwen_association_2015} assumes knowledge of one table to encode the other, and vice versa. Arguably, this approach does not correspond to a practical encoding, like other MDL-based approaches, but also not to a realistic compression scenario, yet it serves as a reasonable motivation for the proposed score.

The proposed score might actually be entirely ad-hoc, in the sense that it does not correspond to the length of an encoding that could be used to represent the data (see for instance \cite{makhalova_numerical_2019} \xinsec{tabular-numerical}).
One might reasonably devise and justify an evaluation measure suited to the problem at hand, but labelling it as following the MDL principle is arbitrary and inappropriate, short of an explanation of how this corresponds to encoding, and can only lead to confusion.

Authors sometimes approach the topic with caution and include disclaimers stating that their proposed methods are inspired by or in the spirit of the MDL principle (see for instance \cite{shokoohi-yekta_discovery_2015} \xinsec{sequential-motifs}). This can be seen as a way to allow oneself to take some liberties with the principle, indeed considering it as a source of inspiration rather than as law, but also as a way to preventively fend off criticism and accusations of heresy.
There is indeed a range of opinions about how closely one must conform to the MDL principle and to information theory.


\subsection{Making comparisons}
\label{sec:discuss-comparisons}
How to make meaningful comparisons between compression-based scores and between corresponding results requires careful consideration.
For instance, one might ponder whether the compression achieved for a dataset is an indication of how much structure is present in it, or at least how much could be detected, and to what extent it can serve as a measure of the performance of the algorithm.

Does it make sense to compare the length of a dataset encoded with the proposed scheme to the original unencoded data? And is it a problem if the latter is shorter? Keeping in mind that compression is used as a tool for comparing models, rather than for practical purposes, we answer both questions in the negative.
The compression achieved with the simplest model, be it the code table containing only elementary patterns such as the singleton itemsets, known as the \emph{standard code table} in \algKrimp{} (\cite{vreeken_krimp_2011} \xinsec{itemsets}), for dictionary-based approaches (cf.\ \citfig{fig:dictionary-binary}\textit{(i)}) or the single-block model for block-based approaches (cf.\ \citfig{fig:blocks-binary}\textit{(i)}), is often considered as a basis for comparison.
The ratio of the compression achieved with a considered model to the compression achieved with the elementary model, known as the \emph{compression ratio}, is then computed and used to compare different models, with lower compression ratios corresponding to better models. This is a way to normalise the scores and allow more meaningful comparisons and evaluations.

\qdone{Models in Fig.2 and in Fig.3 can be compared to each other but not across.}
A direct comparison of the raw description lengths, in terms of numbers of bits, of the same data encoded with different methods is typically not meaningful. For instance, it does not make sense to compare the description lengths reported in \citfig{fig:dictionary-binary} to those reported in \citfig{fig:blocks-binary}. Comparing compression ratios across different methods is not really meaningful either in general. Indeed, an easy way to win this contest would be to design an artificial encoding that penalises very heavily the use of elementary patterns. If the different methods handle compatible pattern languages, comparing the compression ratios achieved when considering as model, in turn, the set of patterns selected by each method and applying either encoding can be of interest, and might shed some light on the respective biases of the methods. If the pattern languages are not compatible, then no quantitative comparison can be devised easily and great care must be taken to choose suitable encodings. Qualitative evaluations of obtained patterns are valuable, despite being subjective and domain dependent. In the end, finding a good set of interesting and interpretable patterns is what matters.  

\noscite{%
\doneedit{condensed the last section into a couple of paragraphs}{%
\subsection{Beyond mining patterns with MDL}
\label{sec:discuss-beyond}

Finally, we highlight two directions of research that do not fall strictly within the realm of MDL-based pattern mining methods, yet are clearly related, constitute recently active and fruitful research topics, and might therefore be of interest to the reader.

The first direction consists of studies of correlation and causality that build on algorithmic information theory in general and, for a few of them, on MDL-based pattern mining techniques more in particular.
For instance, \citet{budhathoki_correlation_2017} propose two algorithmic correlation measures and present practical instanciations based on the MDL principle, using the \algSlim{} and \algname{Pack} algorithms (cf.\ Sections~\ref{sec:itemset-improve} and~\ref{sec:tabular-binary}, respectively). They also propose to use a MDL score to infer the direction of causality between pairs of discrete variables \citep{budhathoki_mdl_2017}.

The second direction consists in the development of pattern mining methods relying on a different modeling approach, also grounded in information theory, namely on maximum entropy modeling.
In particular, \citet{debie_framework_2010} introduced a framework for data mining based on maximum entropy modeling, sometimes referred to as the \emph{FORSIED} framework, for \emph{Formalising Subjective Interestingness}, from which models are derived for different types of assumptions, data and patterns~\citep[including for instance][]{kontonasios_formalizing_2012,kontonasios_maximum_2013,van_leeuwen_subjective_2016,adriaens_subjectively_2019,puolamaki_interactive_2020}.}
}


\secciteTL{%
\begin{srvsection}{Beyond mining patterns with MDL}{beyond}

\oblock{keyword={beyond}, notkeyword={causality}, notkeyword={maxent}, notkeyword={drop}}{%
In this penultimate section, we highlight approaches that do not fall strictly within the category of MDL-based pattern mining methods, yet are clearly related, constitute recently active and fruitful research topics, and might therefore be of interest to the reader. First, we highlight studies of correlation and causality that build on algorithmic information theory in general and, for a few of them, on MDL-based pattern mining techniques more in particular. Second, we outline a framework for pattern mining that relies on a different modeling approach, namely on maximum entropy modeling.
}

\rblock{Correlation and causality}{beyond-causality}{keyword={beyond}, keyword={causality}}{%
A core data analysis problem consists in detecting the presence, measuring the strength, and inferring the direction of dependencies between variables in an observational dataset.
Various methods have been proposed to discover correlated variables and infer the causal structure of a dataset~\citep{pearl_causality_2009}.
In particular, efforts have focused on applying the tools of algorithmic information theory \xcfsec{background-other} to these questions~\citep{janzing_causal_2010}, aiming to increase the scalability of developed methods and reduce their reliance on assumptions about the underlying probabilities and the shape of the relationship linking the variables.

Simply put, looking at how much can be saved by compressing two objects together rather than separately can be used to measure the strength of their correlation. Furthermore, given a pair of objects, comparing how well the first can be compressed given the second, and vice versa, provides an indication about the direction of causality between the objects.
More formally, a central principle in causal inference states that if $x$ causes $y$, it is easier to describe $y$ using $x$ than the other way around~\citep{pearl_causality_2009}.
This principle can be formalised in terms of the Kolmogorov complexity \xcfsec{background}. Specifically, the conditional Kolmogorov complexity of object $x$ given object $y$, denoted $K(x \mid y)$, is the length of the shortest program that generates $x$ and halts, having access to the information in $y$.
Then, if $x$ causes $y$ we expect that there exists a shorter algorithm to describe $y$ given $x$ than the other way around, and hence $K(y \mid x) < K(x \mid y)$.
However, the Kolmogorov complexity is not computable, and various practical instanciations have been proposed, including based on the cumulative and Shannon entropies~\citep{rissanen_measures_1987,vreeken_causal_2015} or on the MDL principle, for instance.

In particular, \citet{budhathoki_correlation_2017} propose two algorithmic correlation measures and present practical instanciations based on the MDL principle, using the \algSlim{} and \algname{Pack} algorithms (cf.\ Sections~\ref{sec:itemset-improve} and~\ref{sec:tabular-binary}, respectively).
\citet{budhathoki_origo_2018} introduce the \algname{Origo} algorithm to infer the direction of causality between binary variables, also relying on the \algname{Pack} algorithm \xcfsec{tabular-binary} to instantiate the MDL score.

Several methods have been proposed to infer the direction of causality between pairs of variables using a MDL score, for discrete variables with refined MDL \citep{budhathoki_mdl_2017}, as well as using classification and regression trees \citep{marx_causal_2019} or global and local regression functions \citep{marx_telling_2019}.
Given a collection of variables $X_1, \dots, X_m$, and $Y$, \citet{kaltenpoth_we_2019} aim to tell whether the $X$ variables jointly cause $Y$, or whether there is an unobserved confounding variable, the real parent, using probabilistic principal component analysis (PCA) and a MDL score.
\citet{mian_discovering_2021} present a method for learning causal graphs where all edges are directed, using multivariate regression and a MDL score.  \citet{marx_formally_2022} aim to elucidate the link between MDL-based estimators and the postulate of algorithmic independence of conditionals that underpins this line of approaches.

Other methods have been presented that rely not on MDL but on other information-theoretic scores such as the Shannon entropy or mutual information and aim to infer the direction of causality between pairs of variables \citep{kocaoglu_entropic_2017,budhathoki_accurate_2018}, to detect functional dependencies between variables \citep{mandros_discovering_2020, pennerath_discovering_2020}, as well as to detect correlations between subspaces \citep{nguyen_universal_2016}, 
to discover causal rules from observational data \citep{budhathoki_rule_2018} or to detect correlations between categorical variables without assumptions on the distribution \citep{mandros_discovering_2019}.
Considering temporal data and Granger causality, \citet{budhathoki_causal_2018} aim to infer the direction of causality between two event sequences using a sequential normalised maximal likelihood (NML) score, while \citet{hlavackova-schindler_poisson_2020} aim to detect causality between timeseries that follow a Poisson distribution, using graphical models and a minimum message length (MML) criterion.
}

\rblock{Maximum entropy modeling}{beyond-maxent}{keyword={beyond}, keyword={maxent}}{%
Considering the core task of mining itemsets and association rules, it became quickly obvious that finding items that frequently co-occur is not enough, and that one needs to consider statistical dependencies between the items (see for instance \cite{silverstein_beyond_1998} \xinsec{background-dm}).
In particular, more insight can be obtained by estimating the expected frequency of items co-occurrence and comparing it to the observed frequency.
Various probabilistic models can be used for the estimation \citep{pavlov_beyond_2003}, including the maximum entropy distribution.

Several approaches are proposed that rely on the maximum entropy distribution to estimate the occurrence frequency of itemsets based on their subsets and contrast this estimate to filter them \citep{meo_theory_2000,jaroszewicz_pruning_2002,wang_summarizing_2006,tatti_maximum_2008}.
Statistics of the dataset, such as marginal counts, can also be used to constrain the model \citep{tatti_using_2010}.

Going beyond local models, \citet{mampaey_tell_2011} define maximum entropy models for the dataset, which allows them to iteratively select a collection of itemsets that summarises the data well. They implement their approach as the \algname{MTV} algorithm (see also Section~\ref{sec:tabular-binary}).
\citet{dalleiger_explainable_2020} look for a collection of patterns and partition of the transactions into components, such that patterns  might be relevant only to a subset of components. The actual mining is done by alternating between two algorithms; \algname{DISC} refines the assignment of transactions to components given a collection of patterns, whereas \algname{DESC} discovers patterns given a partitioning of the data. The latter is essentially a improved variant of the \algname{MTV} algorithm as it optimises the same score but can additionally deal with different data components.

\medskip
Simply put, maximum entropy modeling for pattern mining works as follows. Given some properties of the dataset, a probability distribution is computed over datasets possessing these properties in expectation. The maximum entropy distribution is chosen because this distribution makes no additional assumptions beyond the considered properties and is therefore the least biased. The probability of observing each of the different candidate patterns under this distribution, that is, the probability that the pattern occurs in a dataset with the considered properties, is then evaluated. The lower this probability, the more unexpected and surprising the pattern is considered to be, and hence the more interesting it is deemed. Selected patterns can be seen as discovered properties of the dataset. They can be incorporated as constraints and the probability distribution updated, thereby supporting an iterative, potentially interactive, mining process.

Whereas when following the MDL principle we aim to describe the whole dataset as compactly as possible, the goal when using maximum entropy models is to select the most informative patterns. Typically, selecting all non-redundant patterns that convey information would still produce a large output. Therefore, a criterion must be used to decide when to stop, putting in balance the information content of the patterns and the model complexity.

The constraints imposed on the distribution might capture measured properties of the dataset at hand, but might also reflect the expertise and (possibly incorrect) assumptions of the analyst with respect to the data. The evaluation of the patterns is thus designed to take into account the current experience and understanding of the analyst, albeit in a limited manner. For this reason, the resulting interestingness measure is often called ``subjective''.

\citet{debie_framework_2010} \citep[also][]{debie_information_2011,de_bie_subjective_2013} introduce a framework for data mining based on maximum entropy modeling, sometimes referred to as the \emph{FORSIED} framework, for \emph{Formalising Subjective Interestingness}. They start with the task of mining tiles from a binary database considering assumptions on the row and column marginals \citep{kontonasios_information-theoretic_2010}, then derive models for different types of assumptions~\citep{kontonasios_formalizing_2012}, data and patterns, such as real-valued tabular data \citep{kontonasios_maximum_2013} and various kinds of subgraphs~\citep{van_leeuwen_subjective_2016,adriaens_subjectively_2019,deng_explainable_2020,kapoor_discovering_2020,kapoor_online_2021}, also in a visual interactive exploratory setting~\citep{puolamaki_interactive_2020}. 
\qdone{refs to subgraph patterns}

An intuitive difference between the two families of approaches is that, following the MDL principle, what is most frequent, most expected, results in the most efficient compression. Instead, in maximum entropy modeling, what is most unexpected, deviates most from assumptions, is generally considered most interesting. However, going too far in either direction can be dangerous. Conforming too much to expectations can lead to rather boring results, while very unexpected results can be startling and difficult to interpret.

Choosing the type of patterns of interest and designing the encoding allows to incorporate background information by favouring some patterns over others, yet this is somewhat implicit, indirect and static.
Maximum entropy approaches instead require to model assumptions about the data more explicitly. They tend to be fairly computationally intensive, though much less so than randomisation approaches (see for instance \cite{hanhijarvi_tell_2009} \xinsec{background-dm}), that need to explicitly generate, and possibly mine, a large number of randomised copies of the dataset to achieve comparably precise evaluation. Yet, as with randomisation approaches, formulating anything but simple assumptions about the distribution can be difficult.
On the other hand, unlike MDL-based approaches, most methods relying on the maximum entropy distribution naturally allow for updates, incorporating feedback, and support interactive analysis.
That is, in theory, maximum entropy models allow to incorporate diverse background constraints, in a flexible and potentially interactive manner. However, in practise, this is limited by the fact that constraints can quickly render the optimisation unfeasible.
As a step towards alleviating this limitation, \citet{dalleiger_relaxed_2020} propose an algorithm that dynamically factorises the joint distribution in order to effectively and efficiently approximate the maximum entropy distribution.
}
\end{srvsection}

}

\section{Conclusion}
\label{sec:conclusion}
\qdone{Added few lines of conclusion}

After giving an outline of relevant concepts from information theory and coding, \seccite{and an aper\c{c}u of related theoretical and conceptual contributions, }we reviewed MDL-based methods for mining various types of data and patterns. In particular, we focused on aspects related to the design of an encoding scheme, rather than on algorithmic issues for instance, since the former constitutes the most distinctive ingredient of MDL methodologies, but also a major stumbling block and source of contention.
We pointed out two main strategies that underpin the majority of approaches and that can be used to categorise them. Namely, we distinguished dictionary-based approaches from block-based approaches.
Then, we considered some discussion points pertaining to the use of MDL in pattern mining, and highlighted related problems that constitute promising directions for future research. Indeed, there is still room for further development in mining patterns with MDL-inspired methods, and beyond.


\section*{Acknowledgments}
The author is grateful to Peggy Cellier for her feedback during the preparation and revisions of the manuscript, to Hugo M.\ Proen\c{c}a and Jilles Vreeken for their comments on the first version of this document, and to anonymous reviewers for their comments on later versions of this document. 
The contents of this survey reflect the understanding of the author, any mistakes and misinterpretations are her own.

\vfill
\noindent
Constructive comments as well as pointers to missing related works are most welcome and will be considered to prepare a revision of this survey.

\newpage

\renewbibmacro{lblek}{%
\addspace (Section \ref{ref:\thefield{entrykey}})} 


\newrefcontext{fin}
  \printbibliography[notkeyword={drop}]{}
\end{refsection}


\typeout{get arXiv to do 4 passes: Label(s) may have changed. Rerun}
\end{document}